\documentclass[prd,twocolumn,showpacs,nofootinbib]{revtex4-1}

\usepackage{amsfonts}
\usepackage{amsmath}
\usepackage{amssymb}
\usepackage{bm}
\usepackage{dcolumn}
\usepackage{graphicx}
\usepackage{graphics}
\usepackage[latin1]{inputenc}
\usepackage{latexsym}
\usepackage{rotating}
\usepackage[colorlinks=true]{hyperref}
\usepackage[all]{hypcap} 
\usepackage{xspace} 
\usepackage[usenames]{color}
\usepackage{mathrsfs}

\usepackage{ulem}
\definecolor {darkgreen}{rgb}{0.2,0.7,0.2}

\newcommand\be{\begin{equation}}
\newcommand\ba{\begin{eqnarray}}
\newcommand\ee{\end{equation}}
\newcommand\ea{\end{eqnarray}}

\newcommand\bw{\begin{widetext}}
\newcommand\ew{\end{widetext}}

\newcommand{\nn}{\nonumber}

\newcommand{\GR}{{\mbox{\tiny GR}}}

\newcommand{\MAT}{{\mbox{\tiny mat}}}

\newcommand{\CS}{{\mbox{\tiny CS}}}

\newcommand{\BL}{{\mbox{\tiny BL}}}
\newcommand{\HH}{{\mbox{\tiny H}}}
\newcommand{\HBL}{{\mbox{\tiny H,BL}}}
\newcommand{\LE}{{\mbox{\tiny LE}}}

\newcommand{\ext}{\mathrm{ext}}
\newcommand{\inter}{\mathrm{int}}

\newcommand{\mrm}{\mathrm}

\begin{document}
\title{I-Love-Q Anisotropically: Universal Relations for Compact Stars \\ with Scalar Pressure Anisotropy} 

\author{Kent Yagi}
\affiliation{Department of Physics, Montana State University, Bozeman, MT 59717, USA.}

\author{Nicol\'as Yunes}
\affiliation{Department of Physics, Montana State University, Bozeman, MT 59717, USA.}

\date{\today}

\begin{abstract} 

Certain physical quantities that characterize neutron stars and quark stars (e.g.~their mass, spin angular momentum and quadrupole moment), have recently been found to be interrelated in a manner that is approximately insensitive to their internal structure.
%
Such \emph{approximately universal relations} are useful to break degeneracies in data analysis and model selection for future radio, X-ray and gravitational wave observations. 
%
Although the pressure inside compact stars is most likely nearly isotropic, certain scenarios have been put forth that suggest otherwise, for example due to magnetic fields or phase transitions in their interior. 
%
We here investigate whether pressure anisotropy affects the approximate universal relations, and if so, whether it prevents their use in future astrophysical observations.
%
We achieve this by numerically constructing slowly-rotating and tidally-deformed, anisotropic, compact stars in General Relativity to third order in stellar rotation relative to the mass shedding limit.
%
We adopt simple models for pressure anisotropy where the matter stress energy tensor is diagonal for a spherically symmetric spacetime but the tangential pressure differs from the radial one.
%
We find that the equation of state variation increases as one increases the amount of anisotropy, but within the anisotropy range studied in this paper (motivated from anisotropy due to crystallization of the core and pion condensation),
anisotropy affects the universal relations only weakly. The relations become less universal by a factor of 1.5--3 relative to the isotropic case when anisotropy is \emph{maximal}, but even then they remain approximately universal to $10\%$. 
%
We find evidence that this increase in variability is strongly correlated to an increase in the eccentricity variation of isodensity contours, which provides further support for the emergent approximate symmetry explanation of universality. 
%
Whether one can use universal relations in actual observations ultimately depends on the currently-unknown amount of anisotropy inside stars, but within the range studied in this paper, anisotropy does not prevent the use of universal relations in gravitational wave astrophysics or in experimental relativity.  
%
We provide an explicit example of the latter by simulating a binary pulsar/gravitational wave test of dynamical Chern-Simons gravity with anisotropic neutron stars.
%
The increase in variability of the universal relations due to pressure anisotropy could affect their use in future X-ray observations of hot spots on rotating compact stars.
%
Given expected observational uncertainties, however, the relations remain sufficiently universal for use in such observations if the anisotropic modifications to the moment of inertia and the quadrupole moment are less than 10\% of their isotropic values.

\end{abstract}

\pacs{04.30.Db,04.50Kd,04.25.Nx,97.60.Jd}


\maketitle

\section{Introduction}
\label{sec:intro}

Compact relativistic stars, such as neutron stars (NSs) and quark stars (QSs), are excellent testbeds to probe nuclear and gravitational physics. Their extreme internal density and strong gravity provide access to a regime that is far-removed from that attainable with ground-based laboratories. In particular, observations of compact stars may soon provide information about one of the areas with largest uncertainties in nuclear physics: the equation of state (EoS) (i.e.~the relation between pressure and density) at nuclear and supranuclear densities. This could be achieved through the independent measurement of their mass and radius~\cite{steiner-lattimer-brown,ozel-baym-guver,ozel-review,guver,Lattimer:2013hma,Lattimer:2014sga}, since the mass-radius relation depends sensitively on the EoS~\cite{lattimer_prakash2001,lattimer-prakash-review,Lattimer:2012nd}. One way to obtain such independent measurements is by observing the pulse profile produced by hot spots on the surface of compact stars~\cite{Psaltis:2013zja,Lo:2013ava} with an X-ray telescope, like NICER~\cite{2012SPIE.8443E..13G}. This pulse profile, however does not just depend on the star's mass and radius, but also on its moment of inertia, quadrupole moment and higher multipole moments. Degeneracies between these quantities and the mass and radius prevent measuring the latter accurately, unless the degeneracies can be broken.

Compact stars also provide a unique window into the gravitational interaction in extreme gravitational environments. General Relativity (GR) has passed a plethora of tests with flying colors~\cite{TEGP,will-living,stairs}, but these typically involve quasi-stationary environments with weak gravitational fields, relative to those possible in the vicinity of compact stars. Tests of GR with the latter allow us to confirm Einstein's theory in a regime that has been mostly-unexplored: the ``strong-field.'' For example, in certain scalar tensor theories~\cite{Damour:1993hw,Damour:1996ke} sufficiently compact stars can \emph{spontaneously scalarize}, i.e.~develop a non-trivial scalar field anchored to the star, a strong-field modification of GR that has now been stringently constrained with binary pulsar observations~\cite{Freire:2012mg,Wex:2014nva,Berti:2015itd}. Another way to test GR with compact stars is through the measurement of their mass and radius, since the mass-radius relation depends sensitively on the underlying gravitational theory~\cite{yunes-CSNS,alihaimoud-chen,kent-CSNS,pani-NS-EDGB,eling-AE-NS,Yagi:2013ava,Doneva:2013qva,Yazadjiev:2014cza,Staykov:2014mwa,Yazadjiev:2015zia}. Such a test, however, is strongly degenerate with the star's EoS.

One can break these degeneracies when probing nuclear physics or gravitational physics by using relations among \emph{stellar observables} that do not depend sensitively on the EoS. By ``observables,'' we here mean quantities that characterize the exterior gravitational field of the star, such as its mass, radius, moment of inertia and higher multipole moments associated with their shape. The level to which such relations must be EoS-insensitive or \emph{approximately universal} depends on the accuracy of the astrophysical observation in which they are intended to be used. For example, if one wishes to use the pulse profile of hot spots on the surface of NSs to measure their radius to 5\% accuracy, it suffices if the relation between the moment of inertia and the quadrupole moment is EoS-insensitive to at least 5\%. 

The search for approximately universal relations between observables is not new~\cite{1989ApJ...340..426L,1997PhR...280....1P,lattimer_prakash2001,2004Sci...304..536L,2009A&A...502..605H,andersson-kokkotas-1998,benhar2004,tsui-leung,lau,Yagi:2013sva,kiuchi,kyutoku,Kyutoku:2011vz,2012PhRvL.108a1101B,Bernuzzi:2014kca,Takami:2014zpa,Takami:2014tva,AlGendy:2014eua,Chan:2014kua}, but recently a certain set was discovered that is special for two reasons~\cite{I-Love-Q-Science,I-Love-Q-PRD}. First, the EoS-insensitivity of these relations is typically larger than that found in other observables, with EoS-variability at the percent level. Second, the relations involve quantities that enter observables directly and are related to the multipole moments of the exterior gravitational field of the star, such as its moment of inertia ($I$), its spin angular momentum ($J$), its quadrupole moment ($Q$) and its tidal deformability (or Love number) ($\lambda$). The subset of approximately universal relations between $I$, $\lambda$ and $Q$ is commonly referred to as the \emph{I-Love-Q relations}. Although these approximately universal relations were first discovered studying slowly-rotating and unmagnetized NSs with barotropic and isotropic EoSs~\cite{I-Love-Q-Science,I-Love-Q-PRD}, they were immediately extended to stars with different EoSs~\cite{lattimer-lim}, large tidal deformation~\cite{maselli}, rapid rotation~\cite{doneva-rapid,Pappas:2013naa,Chakrabarti:2013tca,Yagi:2014bxa}, strong magnetic fields~\cite{I-Love-Q-B}, non-barotropic EoSs~\cite{Martinon:2014uua} and to non-GR theories~\cite{I-Love-Q-Science,I-Love-Q-PRD,Sham:2013cya,Pani:2014jra,Doneva:2014faa,Kleihaus:2014lba}. The possible origins of such universality were studied in~\cite{Yagi:2014qua,Sham:2014kea}, with a purely analytic approach developed in~\cite{Stein:2014wpa,Chatziioannou:2014tha,Chan:2014tva}. 

\begin{figure*}[htb]
\begin{center}
\includegraphics[width=8.5cm,clip=true]{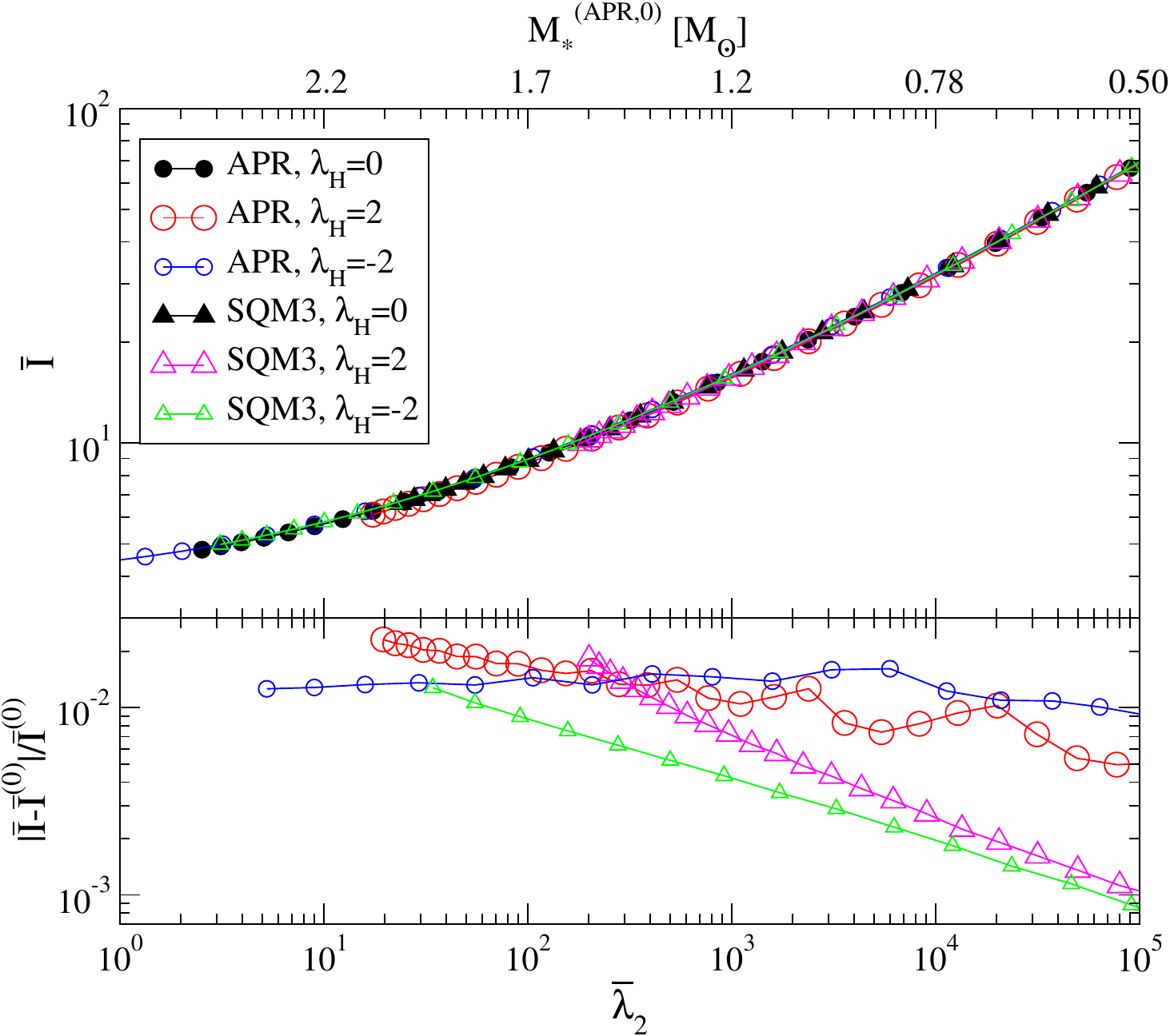}  
\includegraphics[width=8.5cm,clip=true]{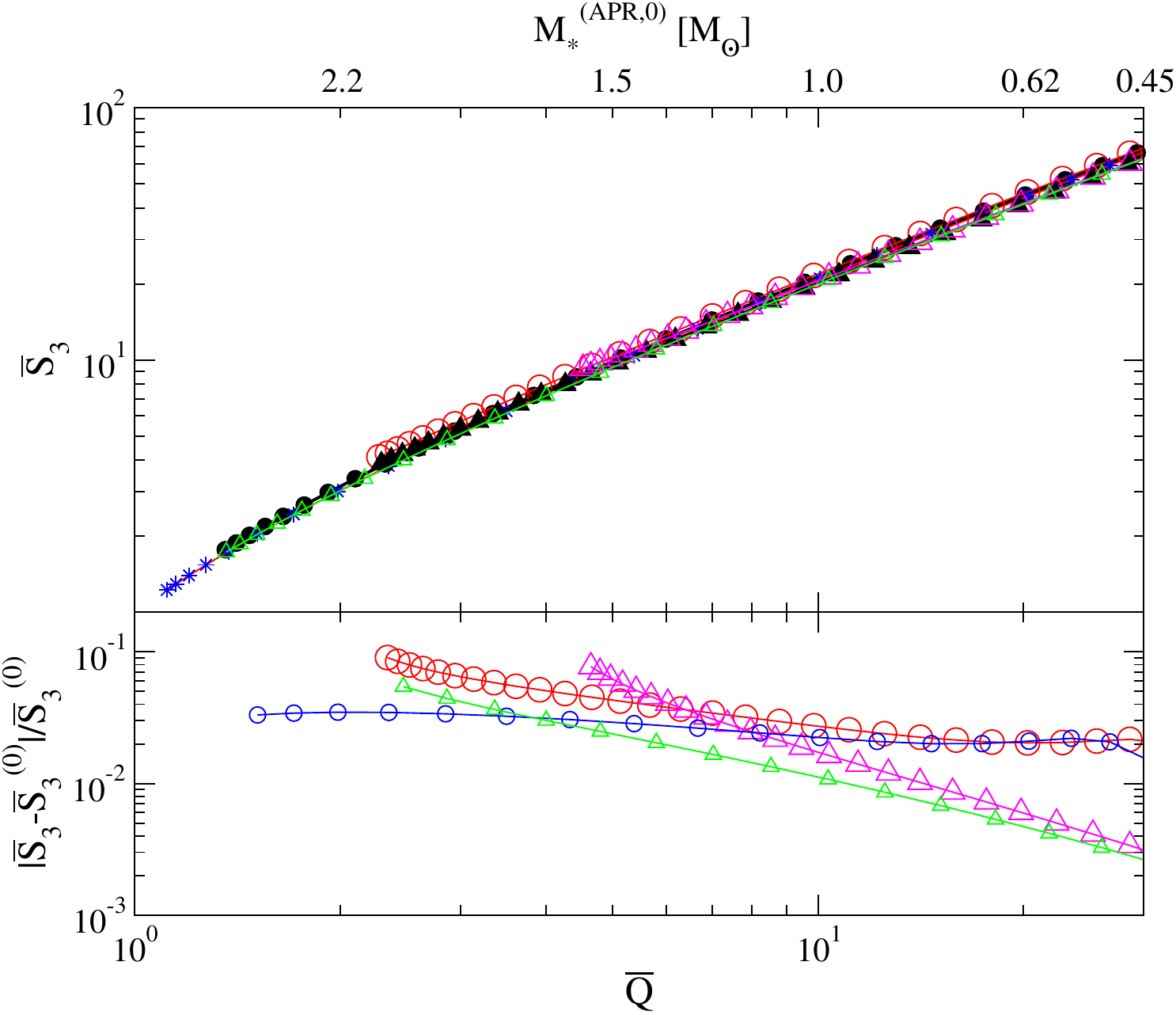}  
\caption{\label{fig:univ-APR-SQM3-intro} (Color online) (Top panels) Relations between the reduced moment of inertia $\bar I = I/M_*^{3}$ and the reduced tidal deformability $\bar \lambda_2 = \lambda_2/M_*^{5}$ (left) and between the reduced quadrupole moment $\bar Q = - Q/(M_*^{3} \chi^{2})$ and the reduced octupole moment $\bar S_3 = - S_{3}/(M_*^{4} \chi^{3}) $ (right) for a sequence of NSs (with an APR EoS) and QSs (with a SQM3 EoS) with various anisotropy parameters $\lambda_\HH$. The quantities used to adimensionalize these observables are $M_*$, the mass of the compact star in the non-rotating limit, and $\chi = J/M_*^{2}$, with $J$ the magnitude of the spin angular momentum. The single parameter along the sequence is the stellar mass (or compactness). For reference, we show the NS mass with an APR EoS and isotropic pressure $M_*^{\mrm{(APR,0)}}$ on the top axis, where the superscript 0 refers to $\lambda_\HH = 0$. (Bottom panels) Fractional relative difference between the relations for anisotropic stars and isotropic stars. Observe that the relations are at most affected to $\sim 2\%$ in the I-Love case and $\sim 10\%$ in the $S_{3}$-Q case due to anisotropy.
}
\end{center}
\end{figure*}

The I-Love-Q relations have direct applications to nuclear physics, experimental relativity and gravitational wave physics. For example, the relation between $I$ and $Q$ can be used to eliminate the I-Q degeneracy in the X-ray pulse profile of hot spots on the surface of compact stars~\cite{Psaltis:2013fha,Baubock:2013gna}. Another example is the use of the relation between $I$ and $\lambda$ to carry out GR tests with combined binary pulsar and GW observations, irrespective of the EoS. Such tests could be particularly powerful, for example allowing constraints that are six orders of magnitude stronger than current Solar System and table-top experiment bounds~\cite{I-Love-Q-Science,I-Love-Q-PRD} on certain quadratic gravity theories~\cite{jackiw,CSreview}. One can also use the relation between $Q$ and $\lambda$ to break the Q-J degeneracy in the observation of GWs emitted during the late, quasi-circular, inspiral of NS binaries~\cite{I-Love-Q-Science,I-Love-Q-PRD}.

Because of their multi-faceted nature and wide applicability to different branches of physics and astrophysics, it is important to determine all possible physical processes that may render the I-Love-Q relations more EoS-sensitive, and thus, less universal. One such physical process is \emph{pressure anisotropy} in the interior of compact stars, i.e.~the interior pressure in the radial direction being different from that in the polar or azimuthal directions. Do we expect NSs to be \emph{strongly anisotropic} and have a large anisotropic pressure component? Several studies exist (see~\cite{1997PhR...286...53H} for a review) that suggest the existence of several sources of anisotropy, such as stellar solid or superfluid cores ~\cite{1990sse..book.....K,1996csnp.book.....G,Heiselberg:1999mq}, relativistic nuclear interactions~\cite{1972ARA&A..10..427R,1974ARA&A..12..167C}, strong magnetic fields~\cite{Yazadjiev:2011ks,1995A&A...301..757B,Konno:1999zv,2001ApJ...554..322C,Ioka:2003nh,2010MNRAS.406.2540C,2012MNRAS.427.3406F,2013MNRAS.435L..43C,2014MNRAS.439.3541P,2015MNRAS.447.3278B}, pion condensation~\cite{Sawyer:1972cq}, phase transitions~\cite{Carter:1998rn} or crystallization of the core~\cite{Nelmes:2012uf}. In fact, relatively simple, two-fluid models with normal and superfluid components are mathematically equivalent to a single anisotropic fluid~\cite{1980PhRvD..22..807L,1997PhR...286...53H}. 

But even if there were a source for anisotropy inside compact stars, one may expect isotropy to be eventually restored. In fact, any restoring force, such as gravity in the strong field, is expected to pull back the compact stellar structure to its isotropic state on the fluid's internal time scale. Given this, it is not clear that NSs should have anisotropic pressure, and if they somehow do, how anisotropic they should be and what mathematical model best describes it. We will here put our personal bias aside and take an agnostic view when asking the following questions\footnote{These questions were already addressed in~\cite{I-Love-Q-B} for a particular example where a large amount of anisotropy is produced by very strong magnetic fields.}:
\begin{enumerate}
\item How does pressure anisotropy affect the I-Love-Q and other universal relations in NSs and QSs?
\item Does pressure anisotropy spoil the use of such relations to break degeneracies in various observations?
\end{enumerate}

We answer these questions by numerically constructing slowly-rotating and tidally-deformed NSs and QSs to third order in a perturbative expansion in $\chi$, a dimensionless parameter constructed from the ratio of the star's spin angular momentum to its mass squared. We assume a scalar anisotropy model for simplicity, where the matter stress energy tensor for a spherically symmetric background is assumed diagonal and the amount of anisotropy is captured by a single parameter. Slowly-rotating, anisotropic NSs had already been constructed to linear order in $\chi$ in GR~\cite{Bayin:1982vw,Silva:2014fca} and in scalar-tensor theories~\cite{Silva:2014fca}, but we here extend these calculations to third order. We model pressure anisotropy through the simple and phenomenological scheme of Horvat \textit{et al.}~\cite{Horvat:2010xf} (the H model), as well as the scheme of Bowers and Liang~\cite{1974ApJ...188..657B} (the BL model). Although the latter is unphysical, i.e.~the pressure anisotropy does not vanish in the non-relativistic limit\footnote{The non-relativistic limit or the \emph{Newtonian limit} here means the leading-order expansion in the gravitational field strength at the stellar surface, or equivalently an expansion in the stellar compactness (the ratio of the stellar mass to the stellar radius).}, it is still useful to gain a better analytical understanding of the physical scenario, which can then be applied to numerical results obtained in the H model.

\subsection{Executive Summary}

Let us first focus on the effect of pressure anisotropy on the universal relations [question (1)]. Following~\cite{Doneva:2012rd,Silva:2014fca}, we consider anisotropic stars in the H model with the anisotropic parameter $\lambda_\HH \in (-2,2)$, where the isotropic limit is recovered when $\lambda_{\HH} = 0$.  Such a choice is mainly motivated from anisotropy due to crystallization of the core~\cite{Nelmes:2012uf} and pion condensation~\cite{Sawyer:1972cq}. Table~\ref{table-summary} shows the maximum variability on different relations when one keeps the EoS fixed and varies $\lambda_{\HH}$ (first row) and when one keeps $\lambda_{\HH}$ fixed to its largest value and varies the EoS. For comparison, the third row shows the EoS variability for isotropic stars~\cite{I-Love-Q-Science,I-Love-Q-PRD,Yagi:2014bxa}, restricted to the same EoSs and stable compact stars considered in this paper, and working to leading order in spin. 
Observe first that the maximum amount of anisotropy increases the level of EoS variability in all relations by a factor of $1.5$--$3$. Nonetheless, all relations considered remain EoS universal to better than 10\%. 
Observe, however, that when the EoS is fixed, the anisotropic variability of the relations is slightly larger. Thus, the \emph{total} variability (EoS plus anisotropy) of the relations is roughly a factor of 2--4 larger than the EoS-variability in the isotropic case. 
One must then conclude that, for the range of anisotropy studied here,
the approximately universal relations remain approximately universal, but to a lesser degree than for isotropic stars. 
{\renewcommand{\arraystretch}{1.2}
\begin{table}[h]
\begin{centering}
\begin{tabular}{c|c|c|c|c}
\hline
\hline
\noalign{\smallskip}
Maximal Variability & I-Q &  I-Love &  Q-Love &  $S_{3}$-Q  \\
\hline
with $\lambda_{\HH}$ (fixed APR EoS) & 7 \% & 2 \% & 8 \%& 9 \% \\
with EoS (fixed $\lambda_{\HH} = 2$) & 5 \% & 2 \% & 5 \% & 8 \%  \\
with EoS (fixed $\lambda_{\HH} = 0$) & 2 \% & 0.7 \% & 2 \% & 5 \% \\
\noalign{\smallskip}
\hline
\hline
\end{tabular}
\end{centering}
\caption{Maximum effect of anisotropy for fixed (APR) EoS (first row), effect of EoS variability for fixed, maximal ($\lambda_{\HH} = 2$) anisotropy (second row) and for vanishing anisotropy (third row) on various approximately universal relations (with $S_3$ representing the current octupole moment). Observe that maximal anisotropy increases the level of variability by a factor of 2--4. 
}
\label{table-summary}
\end{table}
}

Figure~\ref{fig:univ-APR-SQM3-intro} shows the effect of anisotropy in the I-Love (left panel) and $S_{3}$-Q (right panel) relations in more detail, with $S_3$ representing the current octupole moment. The top panels show the relations themselves, with the EoS fixed to APR~\cite{APR} for NSs and SQM3~\cite{SQM} for QSs, while the bottom panels show the fractional relative difference between each curve and the isotropic curve. In these figures, each point represents a numerical solution to the perturbed Einstein equations for a different value of central density and anisotropic parameter. Thus, as one moves from the right to the left of these panels, one is considering a sequence of increasing mass or compactness. Observe that, on average, anisotropy affects the relations by at most a few \% in the I-Love case and $\sim 10\%$ in the $S_3$-Q case, which confirms Table~\ref{table-summary}. Notice also that the maximum variation is realized at the largest positive value of anisotropy, which is why $\lambda_{H} = 2$ was chosen in Table~\ref{table-summary}. The relative fractional error decreases in the non-relativistic limit (as one considers stars with lower compactness) because anisotropy vanishes in this limit for the model we considered.  

Are the above results robust to other anisotropy models? To answer this question, we carry out a numerical study with the BL model and find that the relations in the latter  are very similar to that in the H model for NSs.
In particular, we semi-analytically re-derive the \emph{Newtonian three-hair relations} for anisotropic stars, i.e.~the relations between the exterior multipole moments of stars as functions of only the mass, the spin angular momentum and the quadrupole moment in the non-relativistic limit. We find that the EoS-variation in such relations is consistent with the $S_{3}$-Q universality discussed above in the low-compactness regime. 

Given the robustness of the effect of anisotropy on the universal relations, we can search for an analytical expression that shows their dependence on the anisotropy parameter. We consider the three-hair relations in the non-relativistic limit and perturb the EoS about that for incompressible matter. The latter is chosen because the equations of stellar structure in the non-relativistic limit possess a closed-form analytic solution for such an EoS. We then derive the perturbed three-hair relations analytically~\cite{Chatziioannou:2014tha} and find that they are not modified by anisotropy for any $\ell$th multipole moment at leading order in the EoS deformation. This shows explicitly that the three-hair relations are weakly affected by anisotropy in the non-relativistic limit. 

But why is it that the approximately universal relations are affected at all by anisotropy? In order to answer this question, we must understand the physical origin of the approximate universality in the isotropic case. Reference~\cite{Yagi:2014qua} suggested that the approximate universality is a consequence of the emergence of an approximate symmetry in the high-compactness regime of stellar configurations. The approximate symmetry in question is an approximate self-similarity in the eccentricity profile of isodensity contours. We investigated such contours in the case of anisotropic stars and found that, indeed, the eccentricity variation increases to roughly $\sim 40\%$ inside the star. Such a variation is larger than that for isotropic stars by a factor of four, which explains why the EoS-variation becomes larger for anisotropic stars in the three-hair like relations.

The effect of strong anisotropy on the I-Love-Q relations, however, does not necessarily spoil its applications to various observations [question 2 above]. This is because observations have an intrinsic uncertainty level, so as long as the anisotropy variability is below this level, the I-Love-Q relations can still be used. 
Obviously, the previous statement depends on the amount of anisotropy inside neutron stars, a highly uncertain quantity; however, within the range of anisotropy considered in this paper, the universal relations can still be used
for GW observations because the uncertainty in the measurements of the Love number and the individual spins of compact stars are much greater (order $50\%$ and $10\%$ respectively) than the anisotropy variability of the Q-Love relation.
Similarly, for strong-field tests of GR, the variation due to anisotropy in the I-Love relation is much smaller than the error bars in the measurements themselves, and thus, this variability does not affect our ability to perform such tests.

\begin{figure}[t]
\begin{center}
\includegraphics[width=8.5cm,clip=true]{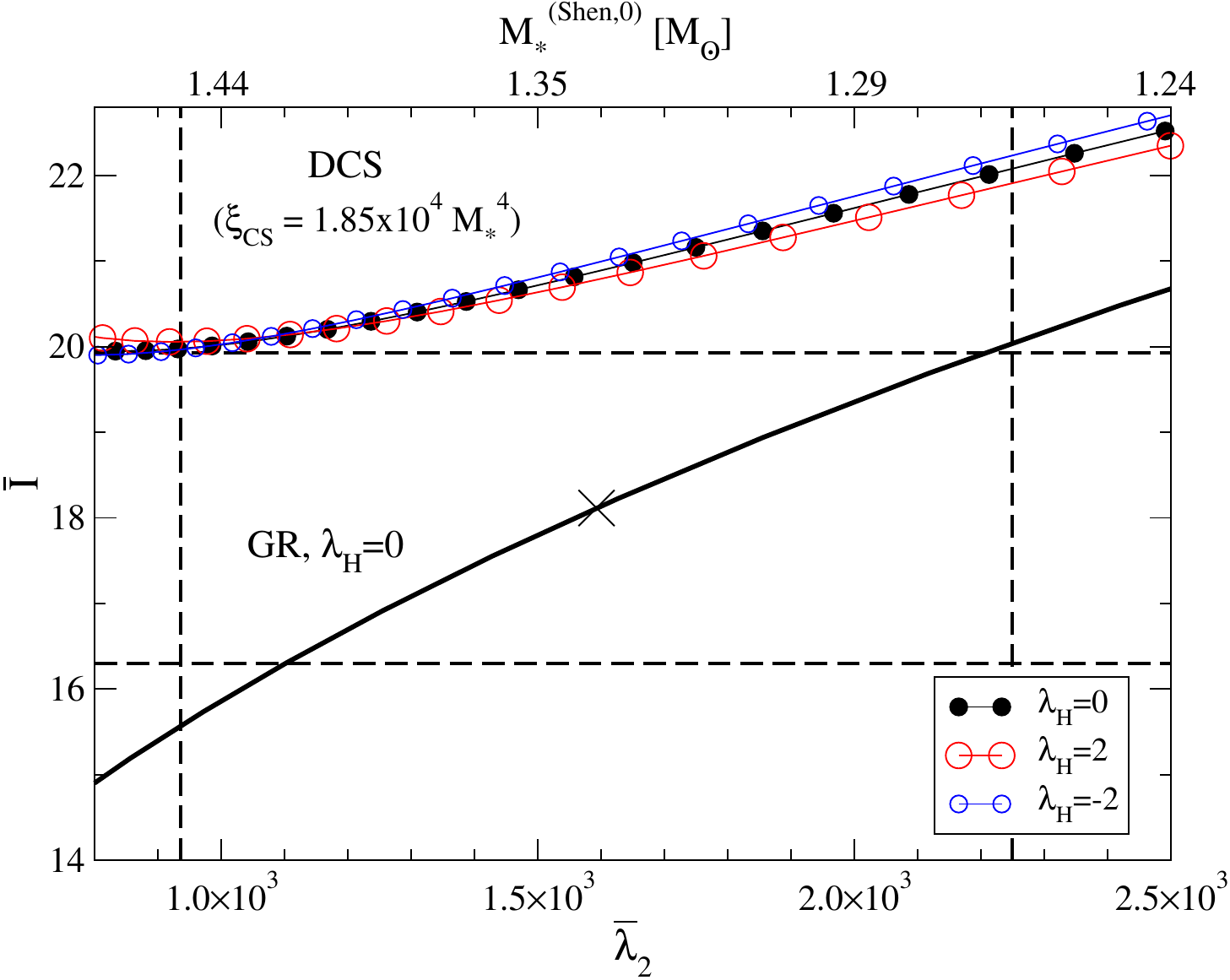}  
\caption{\label{fig:I-Love-Shen-error} (Color online) I-Love relation in GR and in dynamical Chern-Simons gravity~\cite{jackiw,CSreview}. For reference, the mass of an isotropic star with a Shen EoS is shown in solar masses on the top axis. Assume one measures $\bar I$ to 10\% accuracy from a double binary pulsar observation and $\bar \lambda_2$ to 40\% accuracy from a GW observation of a NS binary, with both observations consistent with GR~\cite{I-Love-Q-Science,I-Love-Q-PRD}. These observations are shown with a black cross, while the error ellipses are approximated as an error box shown with black dashed lines. The black curve shows the I-Love relation in GR for isotropic NSs and QSs, while the red, blue and black circles show the I-Love relations in dynamical Chern-Simons gravity for different values of anisotropy parameter and a fixed value of the coupling parameter $\xi_\CS$. Observe that the constraint on $\xi_{\CS}$ remains roughly unaffected by the potential presence of anisotropy.
}
\end{center}
\end{figure}

Let us illustrate the above conclusions with an explicit example. First, we construct slowly-rotating, anisotropic NSs and QSs to linear order in $\chi$ in a particular modified gravity theory, dynamical Chern-Simons gravity~\cite{jackiw,CSreview}. Following~\cite{I-Love-Q-Science,I-Love-Q-PRD}, we assume that we can measure the moment of inertia to $10\%$ accuracy with future double binary pulsar observations and the tidal deformability to $40\%$ with future GW observations. Let us further choose the observations to be for a $1.34 M_\odot$ NS with a Shen EoS, and assume GR is correct. Then, one can draw a fiducial point with an error box around it in the I-Love plane, as shown in Fig.~\ref{fig:I-Love-Shen-error}, where the black curve represents the I-Love relation for isotropic stars in GR. We also show the relations in dynamical Chern-Simons with three different choices of $\lambda_\HH$, but the same value of the coupling constant of the theory. Observe that different values of $\lambda_{\HH}$ hardly affect where the dynamical Chern-Simons I-Love curves cross the error box. Therefore, the constraint that one would derive on the coupling constant of the theory are essentially unaffected by the potential presence of pressure anisotropy in the interior of NSs.

The situation is slightly different for X-ray observations that require the use of the I-Q relation to break degeneracies in the pulse profile model. The goal here is to independently measure the stellar mass and radius to $\sim 5\%$ accuracy, but the variability of the I-Q relation can reach this level for maximal anisotropy and exceed it when varying $\lambda_{\HH}$ for a fixed EoS, as shown in Table~\ref{table-summary}. One can then ask what the maximum value of $\lambda_{\HH}$ is for which the I-Q relation remains universal below $5\%$ when varying over EoSs and anisotropy parameters. The answer to this is $|\lambda_{\HH} | \lesssim 1$, which physically corresponds to a $\leq 10\%$ anisotropy modification to the moment of inertia and quadrupole moment relative to the isotropic case. We then see that the I-Q relations can indeed be used in X-ray observations, even if large anisotropy ($|\lambda_{\HH} | \sim 1$) is allowed in the NS interior. 
 
The rest of the paper is organized as follows.
Section~\ref{sec:formalism} explains the formalism of how to construct slowly-rotating and tidally deformed anisotropic NSs and QSs. 
Section~\ref{sec:num} presents numerical calculations of the universal relations.
Section~\ref{sec:analytic} presents analytic calculations to gain a better physical understanding of the numerical results. 
Section~\ref{sec:applications} explains how the applications of the universal relations to future observations can be affected by anisotropy.
Section~\ref{sec:conclusions} concludes and presents possible avenues for future work. 
Henceforth, we use the geometric units, $c= 1 = G$, throughout. 

\section{Formalism}
\label{sec:formalism}

In this section, we explain how we construct slowly-rotating, compact stars with anisotropic matter to third order in a small spin expansion. We extend following the Hartle-Thorne method~\cite{hartle1967,Hartle:1968ht}, which was invented in the late 1980s to construct slowly-rotating, isotropic NS solutions to quadratic order in $\chi$. This method was extended to third order in~\cite{1976ApJ...207..279Q,1999ApJ...520..788K,benhar} and to fourth order in~\cite{Yagi:2014bxa} for isotropic NSs and QSs. In this section we will explain how this framework can be extended to non-isotropic stars. 

\subsection{Metric Perturbations}

Let us start with by explaining the metric ansatz and introducing the matter stress-energy tensor with anisotropic pressure.
We extend the Hartle-Thorne approach~\cite{hartle1967,Hartle:1968ht}, in which the authors established a formalism to construct slowly-rotating compact stars with isotropic pressure as an expansion to quadratic order in the ratio of the spin angular momentum to the star's mass squared. We use the metric ansatz given by~\cite{Yagi:2014bxa}
\begin{align}
\label{Eq:metric-slow-rot}
ds^2 &= - e^{\nu (r)} \left[ 1 + 2 \epsilon^2 h(r,\theta) \right] dt^2 \nn \\
&+  e^{\lambda (r)} \left[ 1 + \frac{2 \epsilon^2 m(r,\theta)}{r - 2 M(r)} \right] dr^2 
\nn \\
&+  r^2 \left[ 1 + 2 \epsilon^2 k(r,\theta)  \right] 
\left( d\theta^2  + \sin^2 \theta \left\{ d\phi  \right. \right. \nn \\
 & \left. \left.
- \epsilon \left[ \Omega - \omega (r,\theta)  + \epsilon^2 w (r,\theta) \right] dt \right\}^2 \right) + \mathcal{O}(\epsilon^4)\,,
\end{align}
where $\epsilon$ represents a book-keeping parameter that counts the order of slow rotation, $\Omega$ is the stellar angular velocity, $\nu$ and $\lambda$ are the background metric functions while $\omega$, $h$, $k$, $m$ and $w$ are metric perturbations to second and third order in spin. The enclosed mass function $M(r)$ is related to the background metric function $\lambda(r)$ via
\be
e^{- \lambda(r)}  \equiv 1 - \frac{2 M(r)}{r}\,.
\ee

We further decompose the metric perturbations in terms of Legendre polynomials as
\allowdisplaybreaks
\begin{align}
\label{eq:metric-ansatz}
\omega(r,\theta) &= \omega_1(r) P'_1 (\cos \theta)\,, \\
h (r,\theta) &= h_0(r) + h_2(r) P_2(\cos\theta)\,, \\
m (r,\theta) &= m_0(r) + m_2(r) P_2(\cos\theta)\,, \\
k (r,\theta) &= k_2(r) P_2(\cos\theta)\,, \\
w (r,\theta) &= w_{1}(r) P'_1 (\cos \theta) + w_3(r) P'_3 (\cos \theta)\,, 
\end{align}
with $P_\ell' (\cos\theta) = d P_\ell (\cos \theta) /d \cos \theta$. We here used gauge freedom to eliminate the $\ell = 0$ mode in the $k$ function. Following~\cite{hartle1967,Hartle:1968ht}, we introduce a new radial coordinate $R$ such that the stellar density $\rho[r(R,\theta),\theta]$ is the same as the density of the non-rotating configuration $\rho^{(0)} (r)$, namely 
\be
\rho[r(R,\theta),\theta] = \rho(R) = \rho^{(0)}(R)\,.
\ee
In these coordinates, the stellar density does not contain spin perturbation by construction. 
The old and new radial coordinates are related via a new function $\xi(R,\theta)$ by
\be
r(R, \theta) = R + \epsilon^2 \xi(R,\theta) + \mathcal{O}(\epsilon^4)\,.
\ee
As for the other metric perturbations, we decompose $\xi$ through Legendre polynomials:
\be
\label{eq:coord-transf}
\xi(R,\theta) = \xi_0(R) + \xi_{2}(R) P_2 (\cos\theta)\,. 
\ee
In this paper, we are interested in extracting the NS mass $M$, moment of inertia $I$, quadrupole moment $Q$ and octupole moment $S_3$ to leading order in spin, where $I$ is defined by $I \equiv J/\Omega$ with $J$ the magnitude of the spin angular momentum. Therefore, at any $\mathcal{O}(\epsilon^n)$, we only consider the $\ell = n$ mode since that will be the only one that contributes.

\subsection{Matter Representation}

Let us next explain how we choose to represent the matter degrees of freedom.For simplicity, we consider a scalar anisotropy model, where the matter stress energy tensor with anisotropic pressure is given by~\cite{Doneva:2012rd,Silva:2014fca}
\be
\label{eq:matter}
T_{\mu \nu} = \rho u_\mu u_\nu + p k_\mu k_\nu + q \Pi_{\mu \nu}\,.
\ee
Here $u^\mu$ is the fluid four-velocity
\be
u^\mu = (u^0, 0,0,\epsilon \Omega u^0)\,,
\ee
with $u^0$ determined through the normalization condition $u^\mu u_\mu = -1$. In Eq.~\eqref{eq:matter}, $k^\mu$ is a unit radial vector that is spacelike ($k^\mu k_\mu = 1$) and orthogonal to the four-velocity ($k^\mu u_\mu=0$), while $\Pi_{\mu \nu}$ is a projection operator onto a two-surface orthogonal to $u^\mu$ and $k^\mu$:
\be
\Pi_{\mu \nu} = g_{\mu \nu} + u_\mu u_\nu - k_\mu k_\nu\,,
\ee
where $p$ and $q$ are the radial and tangential pressures respectively.
Following~\cite{Doneva:2012rd,Silva:2014fca}, we introduce the anisotropy parameter $\sigma \equiv p-q$, with $\sigma = 0$ corresponding to isotropic matter, and we Legendre decompose it as
\be
\sigma(R,\Theta) = \sigma_0 (R) + \sigma^{(2)}_{0}(R) + \sigma^{(2)}_{2}(R) P_2(\cos\theta) + \mathcal{O}(\epsilon^4)\,.
\ee
We take the radial pressure to be barotropic, $p=p(\rho)$, and hence, its perturbation vanishes from the definition of the new radial coordinate $R$. In a more generic situation, one could introduce pressure anisotropy as a tensor; in that case, the stress energy tensor would have off-diagonal anisotropy components. This is indeed the case when anisotropy is caused by e.g. magnetic fields~\cite{1995A&A...301..757B,Konno:1999zv,2001ApJ...554..322C,Ioka:2003nh,2010MNRAS.406.2540C,2012MNRAS.427.3406F,2013MNRAS.435L..43C,2014MNRAS.439.3541P,2015MNRAS.447.3278B}.

The barotropic radial pressure will be modeled as follows. When considering NSs, we adopt the APR~\cite{APR}, SLy~\cite{SLy}, LS220~\cite{LS}, Shen~\cite{Shen1,Shen2}, WFF1~\cite{Wiringa:1988tp} and ALF2~\cite{Alford:2004pf} EoSs. The first five represents NSs with n,p,e,$\mu$ matter, while the ALF2 EoS represents hybrid stars with n,p,e,$\mu$ and quark matter. We choose these EoSs because the maximum masses they predict for isotropic, non-rotating NSs are all above $2M_\odot$~\cite{2.01NS}. We impose the neutrino-less and $\beta$-equilibrium conditions with a temperature of 0.1MeV and nuclear incompressibility of 220MeV and 281MeV for the LS and Shen EoSs respectively. When considering QSs, we adopt the SQM1, SQM2 and SQM3~\cite{SQM} EoSs based on the MIT bag model. When doing analytic calculations in the Newtonian limit, it will be convenient to also use a polytropic EoS, given by
\be
p = K \rho^{1 + 1/n}\,,
\ee
where $K$ and $n$ are the polytropic constant and the polytropic index respectively. Notice that in the Newtonian limit, the energy density $\rho$ reduces to the rest-mass density.

The anisotropic part of the pressure will be modeled as follows. We will mainly adopt the model proposed by Horvat \textit{et al.}~\cite{Horvat:2010xf} (the H model), which is defined by~\cite{Horvat:2010xf,Doneva:2012rd,Silva:2014fca}
\be
\sigma_0 = 2 \lambda_\HH p \frac{M}{R}\,, 
\ee
for a non-rotating configuration. Here, $\lambda_\HH$ is a parameter that characterizes the amount of anisotropy. Obviously, when $\lambda_\HH=0$, the anisotropy parameter vanishes and one recovers the isotropic case. This model is constructed such that the effect of anisotropy vanishes in the hydrostatic equilibrium equation in the non-relativistic limit, where $p \ll \rho$. $\sigma_0$ vanishes at the stellar center, which is a required boundary condition such that physical quantities, like the mass and spin angular momentum, do not contain singularities~\cite{1974ApJ...188..657B}. Notice that $\sigma_0$ is continuous at the surface even if the density is discontinuous there. 

The H-model is a phenomenological description of anisotropy, where the amount of the latter is controlled by $\lambda_{\HH}$. This parameter can be of order unity when anisotropy is realized by pion condensation~\cite{Sawyer:1972cq}. Alternatively, if the origin of anisotropy is strain in Skyrme crystals~\cite{Nelmes:2012uf},  $\lambda_\HH \approx -2$ for a NS with maximum mass. As the mass decreases, $\lambda_{\HH}$ approaches zero until it vanishes for masses smaller than $1.49M_\odot$. In order to treat both models mentioned above simultaneously, we follow Refs.~\cite{Doneva:2012rd,Silva:2014fca} and consider $\lambda_\HH$ in the range $-2 \leq \lambda_\HH \leq 2$.
Of course, larger degrees of anisotropy are not excluded, but since we currently lack a robust prediction of $\lambda_\HH$, we believe it is natural to adopt the anisotropy range used in previous works~\cite{Doneva:2012rd,Silva:2014fca} as a first step.

In order to carry out analytic calculations and to obtain a better understanding of numerical results, we will also consider an alternative anisotropy model, proposed by Bowers and Liang~\cite{1974ApJ...188..657B} (the BL model). In this model, $\sigma_0$ is given by 
\be
\sigma_0 =  \frac{\lambda_\BL}{3} (\rho + 3p) (\rho + p) \left( 1-\frac{2M}{R} \right)^{-1} R^2\,, 
\ee
for a non-rotating configuration. As in the H model, $\lambda_\BL$ is a constant parameter that quantifies the amount of anisotropy, $\lambda_\BL = 0$ leading to stars with isotropic pressure. The BL model was constructed specifically so that one can solve the modified Tolman-Oppenheimer-Volkoff (TOV) equation for an incompressible anisotropic star analytically~\cite{1974ApJ...188..657B}. Unlike the H model, $\sigma_0$ becomes discontinuous at the surface if the density is also discontinuous. Although the BL model satisfies the boundary condition at the stellar center, the effect of anisotropy in the hydrostatic equilibrium equation does not vanish in the non-relativistic limit. This fact is very unphysical, if one requires pressure anisotropy to be caused by the strain of nuclear matter or magnetic fields at the core of compact stars. Notice also that $\sigma_0$ can be discontinuous at the surface since it depends on the stellar density. On the other hand, the BL model can be useful to track anisotropy calculations analytically. For these reasons, we mainly focus on the H model, and use BL model secondarily for analytic investigations.

\subsection{Field Equations}

Next, we derive the field equations for slowly-rotating anisotropic compact stars.
Substituting Eqs.~\eqref{eq:metric-ansatz} and~\eqref{eq:matter} into the Einstein equations and the equation of motion $\nabla^\mu T_{\mu \nu} = 0$, at $\mathcal{O}(\epsilon^0)$, one finds~\cite{Doneva:2012rd,Silva:2014fca}
\begin{align}
\label{eq:dMdR}
\frac{d M}{dR} &= 4 \pi R^2 \rho\,, \\
\frac{d \nu}{dR} &= 2\frac{4 \pi R^3 p + M}{R(R-2M)}\,, \\
\label{eq:dPdR}
\frac{dp}{dR} &= -\frac{(4\pi R^3 p + M) (\rho + p)}{R(R-2M)} - \frac{2\sigma_0}{R}\,,
\end{align}
while at $\mathcal{O}(\epsilon)$, one finds~\cite{Bayin:1982vw,Silva:2014fca}
\begin{align}
\frac{d^2 \omega_1}{d R^2} &= 4\frac{\pi R^2 (\rho +p) e^{\lambda}-1}{R}\frac{d \omega_1}{d R} \nn \\
 & + 16 \pi (\rho +p-\sigma_0) e^{\lambda} \omega_1\,,
\end{align}
and at $\mathcal{O}(\epsilon^2)$, the $\ell = 2$ mode satisfies
\bw
\begin{align}
\label{eq:sigma2}
\sigma^{(2)}_{2} &=  (\rho + p - \sigma_0) h_2 - \frac{\sigma_0}{R-2M} m_2 + \left[ \frac{(\rho + p) (4 \pi p R^3 + M)}{R (R-2 M)} +\frac{d\sigma_0}{dR} + 2 \frac{\sigma_0}{R} \right] \xi_{2} + \frac{R^2}{3}  (\rho + p - \sigma_0) e^{-\nu}  \omega_1^2\,, \\
\label{eq:m2}
m_2 &= -R e^{-\lambda} h_2  +\frac{1}{6} R^4 e^{-(\nu+\lambda)} \left[ R e^{-\lambda} \left( \frac{d \omega_1}{dR}\right)^2 + 16 \pi R \omega_1^2 (\rho +p-\sigma_0) \right]\,, \\
\label{eq:k2R}
\frac{dk_2}{dR} &= -\frac{dh_2}{dR} + \frac{R-3M-4\pi pR^3}{R^2} e^{\lambda} h_2  + \frac{R-M+4\pi p R^3}{R^3} e^{2\lambda} m_2\,, \\
\label{eq:h2R}
\frac{dh_2}{dR} &= \frac{3 e^{\lambda}}{R} h_2 - \frac{4 \pi p R^3 - M + R}{R} e^{\lambda} \frac{dk_2}{dR} + \frac{2 e^{\lambda}}{R} k_2 + \frac{8  \pi p R^2 + 1} {R^2} e^{2 \lambda} m_2 + \frac{1}{12} R^3 e^{-\nu} \left( \frac{d \omega_1}{dR} \right)^2 \nn \\
& +  4 \pi (\rho+p) e^{2\lambda} \frac{4 \pi p R^3 + M}{R} \xi_{2} + 8 \pi e^{\lambda} \sigma_0 \xi_{2}\,, \\
\label{eq:xi2}
\frac{d\xi_{2}}{dR} &=  \frac{R-2M}{6 R [(\rho + p) (4 \pi p R^3 + M) + 2 (R-2M) \sigma_0]} \left\{- 6 R^2 (\rho + p) \frac{d h_2}{dR} - 12 \sigma_0 R^2 \frac{d   k_2}{dR}\right. \nn \\
  &- \left. 3 \left[ R^2 (\rho + p) \frac{d^2 \nu}{dR^2} -4 \sigma_0 \right] \xi_{2}-12 R \sigma^{(2)}_{2} + 2 R^3  (\rho + p - \sigma_0) e^{-\nu} \omega_1 \left[ \left( R \frac{d \nu}{dR} -2 \right) \omega_1 - 2 R \frac{d \omega_1}{dR} \right]  \right\}\,.
\end{align}
%
Eqs.~\eqref{eq:sigma2}--\eqref{eq:h2R} agree with those for isotropic matter derived in~\cite{hartle1967,Hartle:1968ht} when $\sigma = 0$, while Eq.~\eqref{eq:xi2} becomes linearly dependent in the isotropic case. At $\mathcal{O}(\epsilon^3)$, the $\ell = 3$ mode satisfies
\begin{align}
\frac{d^2 w_3}{dR^2} &= 4 \frac{\pi \rho  R^3 + \pi p R^3 + 2 M - R}{R^2} e^{\lambda} \frac{d w_3}{dR} + 2  \frac{8 \pi \rho R^2 + 8 \pi p R^2 + 5  - 8 \pi R^2  \sigma_0}{R^2} e^{\lambda} w_3 \nn \\
& -    \frac{1}{5} \left( \frac{dh_2}{dR} - 4 \frac{dk_2}{dR} + \frac{e^{\lambda}}{R} \frac{dm_2}{dR} - \frac{1 -8 \rho \pi R^2}{R^2} e^{2\lambda} m_2 \right)\frac{d\omega_1}{dR} - \frac{32}{5} \pi (\rho + p) \frac{e^{\lambda}}{R} m_2 \omega_1 \nn \\
& +   \frac{16}{5} \pi \frac{e^{\lambda}}{R^2} \left[ R^2 \frac{d\rho}{dR} - (4 \pi p R^3 + M) (\rho + p) e^{\lambda}  \right]  \xi_{2} \omega_1 + \frac{32}{15} \pi R^2 (\rho + p) e^{-\nu+\lambda} \omega_1^3 \nn \\
& - \frac{32}{15} \pi \frac{e^{2 \lambda}}{R} \left( R^3  e^{-\nu - \lambda} \omega_1^2 +3 e^{-\lambda} \xi_{2} - 3 m_2 \right) \omega_1 \sigma_0 -\frac{16}{5} \pi \omega_1 e^{\lambda} \xi_{2} \frac{d\sigma_0}{dR} + \frac{16}{5} \pi e^{\lambda} \omega_1 \sigma^{(2)}_{2}\,.
\end{align}
\ew
This equation agrees with that for isotropic matter~\cite{benhar} when $\sigma =0$.

\subsection{Boundary Conditions}

We here explain the boundary conditions that one needs to impose to solve the field equations derived in the previous subsection. We solve the equations from the NS core radius $R=r_\epsilon$ to the surface $R=R_*$ using a fourth order Runge-Kutta algorithm. We set $r_\epsilon = 100$cm and the stellar radius $R_{*}$ is determined from the condition $p(R_*)=0$. The first boundary condition is set at $R=r_\epsilon$, while the second condition is set at the surface. We treat each of these separately below. 

\subsubsection{At the Core Radius}

The boundary condition at the core can be obtained through a local analysis of the solution to the structure equations near the stellar center. One finds
\begin{align}
\allowdisplaybreaks 
\rho (r_\epsilon) &= \rho_c + \rho_2 r_\epsilon^2 + \mathcal{O}(x^3)\,, \\
M(r_\epsilon) &= \frac{4\pi}{3} \rho_c r_\epsilon^3 + \mathcal{O}(x^5)\,, \\
\nu(r_\epsilon) &= \nu_c + \frac{4\pi}{3} (3 p_c+ \rho_c) r_\epsilon^2 + \mathcal{O}(x^4)\,, \\
\omega_1(r_\epsilon) &= \omega_{1c} + \frac{8\pi}{5} (\rho_c + p_c) \omega_{1c} r_\epsilon^2 + \mathcal{O}(x^3)\,, \\
h_2 (r_\epsilon) &=  C_1 r_\epsilon^2 + \mathcal{O}(x^3)\,, \\
k_2(r_\epsilon) &= - C_1 r_\epsilon^2 + \mathcal{O}(x^3)\,, \\
m_2(r_\epsilon) &= - C_1 r_\epsilon^3 + \mathcal{O}(x^4)\,, \\
w_3 (r_\epsilon) &= C_2  r_\epsilon^2 + \mathcal{O}(x^3)\,,
\end{align}
for both the H and BL models, where we have expanded about $x \equiv r_{\epsilon}/R_{*} \ll 1$. Here, $\rho_c$ is the central density, while  $\nu_c$, $\omega_{1c}$, $C_1$ and $C_2$ are constants that are determined by matching the interior and the exterior solutions at the NS surface. $\rho_c$ and $\rho_2$ are obtained through the EoS. On the other hand, the asymptotic behavior of $p$, $\xi_{2}$ and $\sigma^{(2)}_{2}$ at the stellar center are given by
\begin{align}
p^\HH (r_\epsilon) &= p_c - \frac{2\pi}{3}  [(3 p_c + \rho_c) (p_c + \rho_c) + 4 \lambda_H p_c \rho_c] r_\epsilon^2 \nn \\
& +  \mathcal{O}(x^3)\,, \\
\xi_{2}^\HH (r_\epsilon) &= - \frac{(p_c+\rho_c)  (\omega_{1c}^2 e^{-\nu_c}  + 3 C_1)}{4 \pi [(3 p_c + \rho_c) (p_c + \rho_c) + 4 \lambda_H p_c \rho_c]}  r_\epsilon \nn \\
& +  \mathcal{O}(x^2)\,, \\
\sigma^{(2)}_{2}{}^\HH (r_\epsilon) &= - \frac{4  \lambda_\HH (p_c + \rho_c) p_c \rho_c (  \omega_{1c}^2 e^{-\nu_c} + 3 C_1) }{3 [ (p_c + \rho_c) (3 p_c + \rho_c) + 4 \lambda_\HH p_c \rho_c ] } r_\epsilon^2 \nn \\
& +  \mathcal{O}(x^3)\,, 
\end{align}
for the H model and
\begin{align}
p^\BL (r_\epsilon) &= p_c - \frac{1}{3}  (3 p_c + \rho_c) (p_c + \rho_c) (2 \pi +  \lambda_H ) r_\epsilon^2 \nn \\
&+  \mathcal{O}(x^3)\,, \\
\xi_{2}^\BL (r_\epsilon) &= - \frac{ \omega_{1c}^2 e^{-\nu_c} + 3 C_1}{2(3 p_c+ \rho_c) (\lambda_\BL+2 \pi)}  r_\epsilon + \mathcal{O}(x^2)\,, \\
\sigma^{(2)}_{2}{}^\BL (r_\epsilon) &= - \frac{\lambda_\BL}{3 (2 \pi +  \lambda_\BL)} (p_c+\rho_c) \left( \omega_{1c}^2 e^{-\nu_c} + 3C_1 \right) r_\epsilon^2 \nn \\
& +  \mathcal{O}(x^3)\,, 
\end{align}
for the BL model respectively, with $p_c $ representing the central pressure.

\subsubsection{At the Surface}

The boundary conditions at the surface can be obtained by finding the exterior solutions to the structure equations, which then necessitates a local analysis at spatial infinity. Assuming asymptotic flatness at spatial infinity, the exterior solutions are given in~\cite{hartle1967,Hartle:1968ht,benhar,I-Love-Q-PRD,Yagi:2014bxa}. Such solutions, of course, do not depend on whether one considers the H or BL models, since pressure vanishes in the exterior. The exterior solutions contain integration constants that are determined by matching the interior and exterior solutions at the surface. In fact, it is these constants that determine the (Geroch-Hansen) multipole moments of the exterior gravitational field~\cite{Yagi:2014bxa}. For example, the moment of inertia $I$, quadrupole moment $Q$ and octupole moment $S_3$ can be read off from the asymptotic behavior of the exterior solutions for $\omega_1$, $h_2$ and $w_3$ at spatial infinity as~\cite{Yagi:2014bxa}
\begin{align}
\omega_1^\ext (R) &= \Omega \left( 1 - \frac{2 I}{R^3} \right)\,, \\
h_2^\ext (R) &= - \frac{Q}{R^3} + \mathcal{O} \left( \frac{1}{R^4} \right)\,, \\ 
w_3^\ext (R) &= \frac{2 S_3}{3R^5} + \mathcal{O} \left( \frac{1}{R^6} \right)\,.
\end{align}

With the exterior solutions at hand, we can now discuss the boundary conditions at the surface.
Since we are neglecting the NS crust in this paper, i.e.~we are neglecting the fact that the NS
density decays very rapidly at the stellar surface leading effectively to a discontinuity, 
the matching condition for any metric perturbation $A$ at the surface is given by
\be
\label{eq:matching}
A^\inter (R_*) = A^\ext (R_*)\,, \quad  A^\inter {}' (R_*) = A^\ext {}' (R_*)\,,
\ee
where  the superscript ``int'' refers to the interior solutions and $A = \omega_1$, $h_2$ and $w_3$.
On the other hand, if the density is strongly discontinuous at the surface (such as in the case of incompressible NSs or the case of QSs), one needs to be careful with the matching and incorporate the appropriate jump conditions. Following~\cite{damour-nagar,alihaimoud-chen}, we integrate the field equations from $R=R_* - \varepsilon$ to $R=R_* + \varepsilon$ and take the limit of $\varepsilon \to 0$. Then, one finds the jump conditions
\be
\label{eq:matching-jump}
 A^\inter {}' (R_*) +j_A \; \rho_* = A^\ext {}' (R_*)\,,
\ee
with
\begin{align}
j_{\omega_1} &= 0\,, \\
j_{h_2} &= - \frac{4 \pi  R_*  (2 R_*^2 M_* \omega_{1,*}^2   e^{-\nu_*} - 3 \xi_{2,*})}{3 (R_*-2 M_*)}\,, \\
j_{w_3} &=  \frac{8\pi}{15} R_*^4 \omega_{1,*}^2 \omega_{1,*}'  e^{-\nu_*} - \frac{16 \pi R_* \omega_{1,*}  \xi_{2,*} }{5 (R_*-2 M_*)}  \,,
\end{align}
and the subscript $*$ represents the quantity being evaluated at the surface from the interior.

\subsection{Tidally-deformed Compact Stars}

The construction of compact stars with small amounts of tidal deformation proceeds similarly. The field equations that one needs to solve are the same as those for slowly-rotating stars, except that now one must set $\omega_1 = 0$. Therefore, one must also set $\omega_{1c}=0$ in the local analysis of the metric functions near the core. Regarding the exterior solutions, one does not impose asymptotic flatness anymore. This allows us to extract, not only the quadrupole moment $Q$, but also the ($\ell=2$) tidal potential $\mathcal{E}$, whose ratio gives the $\ell = 2$ (electric-type), tidal deformability $\lambda_2$~\cite{hinderer-love,damour-nagar,binnington-poisson}:
\be
\lambda_2 \equiv - \frac{Q}{\mathcal{E}}\,.
\ee
$\lambda_2$ is related to the $\ell = 2$ (electric-type) tidal Love number $k_2$ via $k_2 = (3/2) (\lambda_2/R_*^5)$. The latter can be obtained by matching the interior and exterior solutions at the surface. Equation~(23) in~\cite{hinderer-love} provides an expression for $k_2$ as a function of $y= R_* h'_{2}(R_*)/h_2(R_*)$. For a star with a discontinuous density at the surface, one needs to replace $y$ by $y = R_* h'_{2}(R_*)/h_2(R_*) + 4 \pi  \rho_* R_*^2 \xi_{2,*}/[h_{2,*}(R_*-2M_*)]$.

\section{Universal Relations for Anisotropic Neutron Stars \\ and Quark Stars}
\label{sec:num}

\begin{figure*}[htb]
\begin{center}
\includegraphics[width=8.5cm,clip=true]{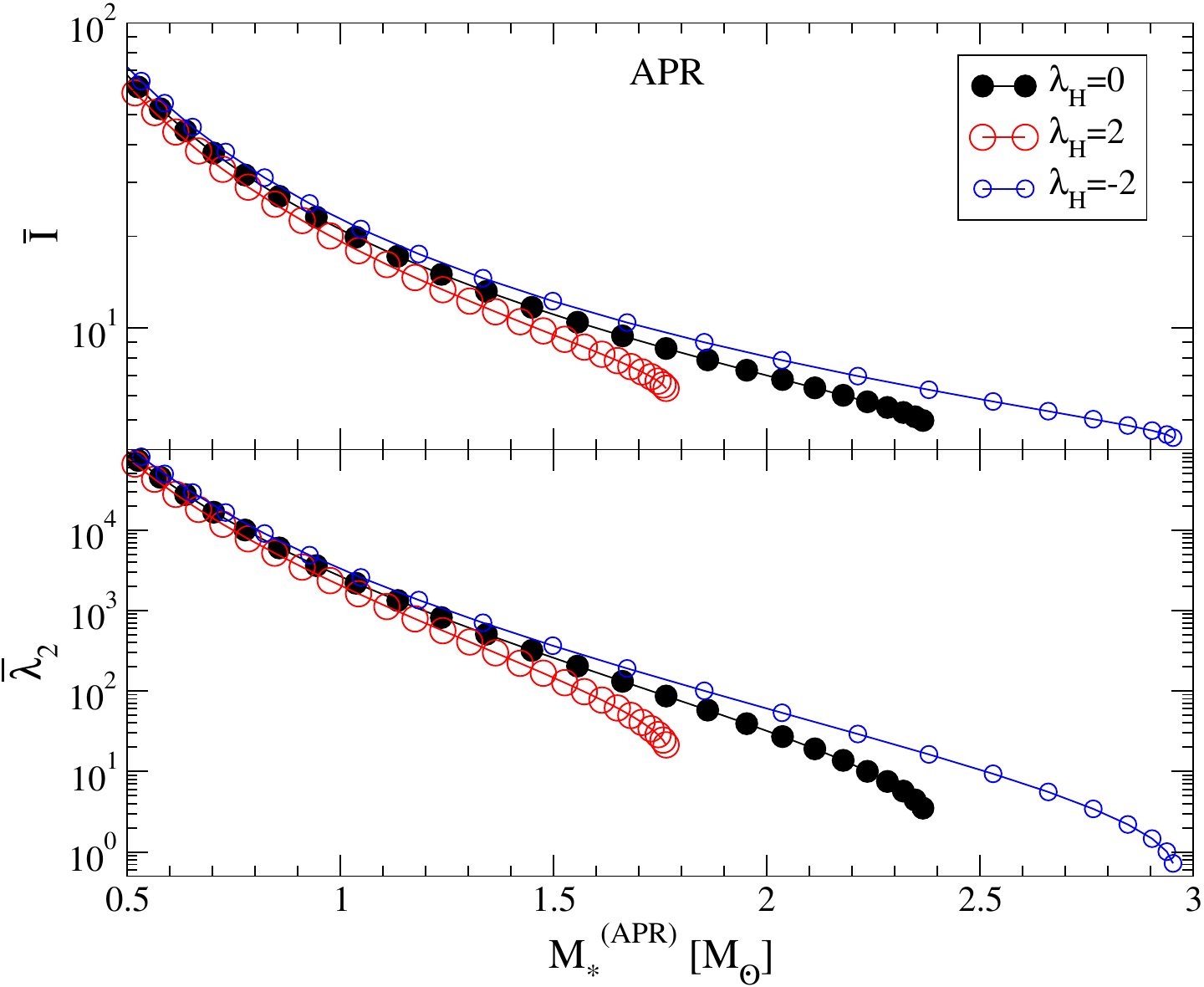}  
\includegraphics[width=8.5cm,clip=true]{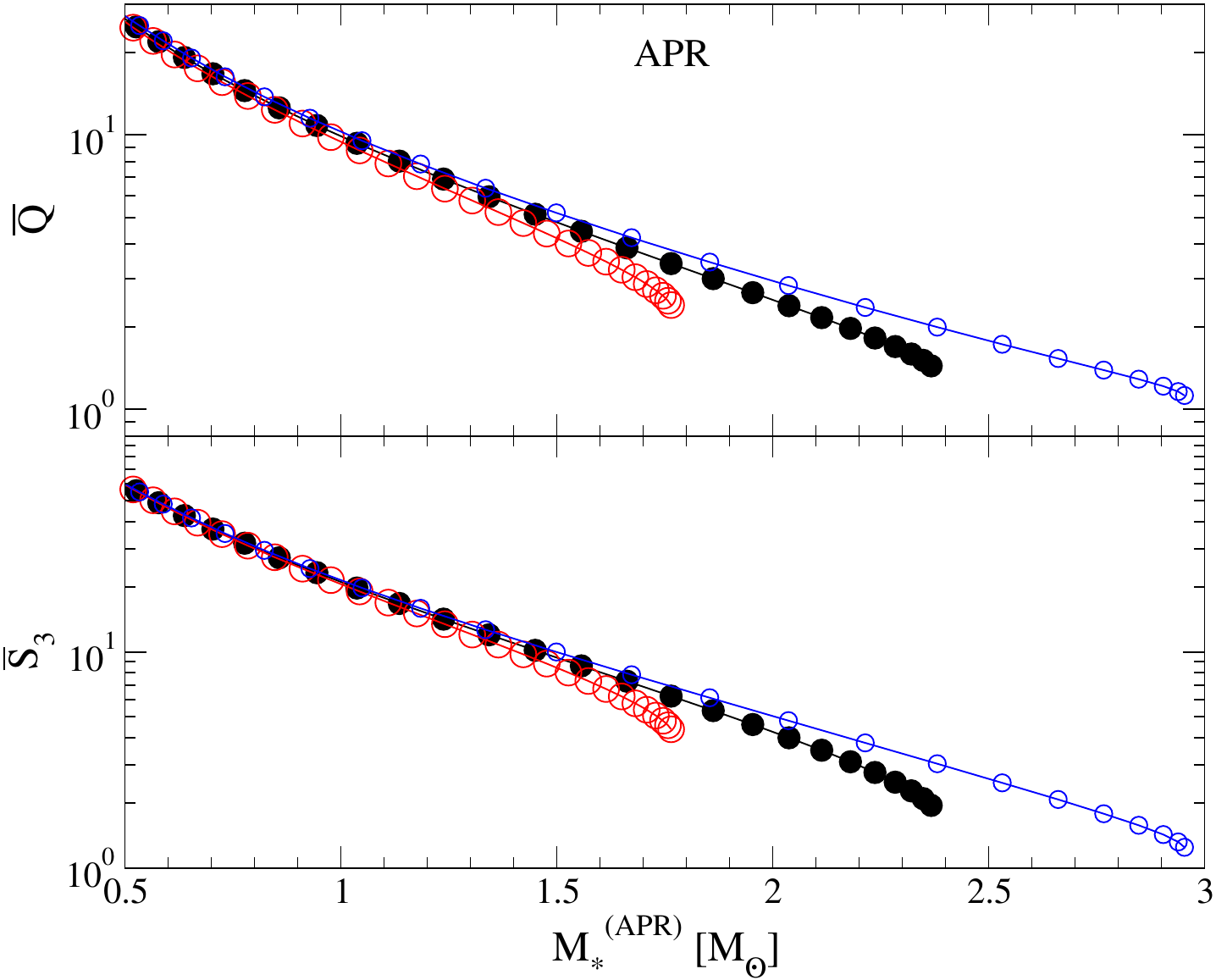}  
\caption{\label{fig:M-dep} (Color online) Mass dependence of $\bar{I}$ (top left), $\bar \lambda_2$ (bottom left), $\bar Q$ (top right) and $ \bar S_3$ (bottom right) for various anisotropy parameters with an APR EoS. Observe that the relations for anisotropic NSs show larger deviations from the isotropic case as one increases the NS mass.}
\end{center}
\end{figure*}

We now present the main numerical results of this paper and try to address the first question posed in Sec.~\ref{sec:intro}, namely, how pressure anisotropy affects the approximately universal relations of compact stars. We look at the relations among the following dimensionless quantities:
\be
\bar I \equiv \frac{I}{M_*^3}\,, \quad \bar Q \equiv - \frac{Q}{M_*^3 \chi^2}\,, \quad S_3 \equiv - \frac{S_3}{M_*^4 \chi^3}\,, \quad \bar \lambda_2 \equiv \frac{\lambda_2}{M_*^5}\,,
\ee
where $\chi \equiv J/M_*^2$ with $M_*$ and $J$ representing the stellar mass (of the non-rotating configuration) and the magnitude of the spin angular momentum respectively.

Before presenting the relations among the dimensionless observables above, let us first look at the dependence of each observable on the stellar mass, which we present in Fig.~\ref{fig:M-dep} for $\lambda_\HH = 2,0,-2$ with an APR EoS. Observe that the deviations induced by pressure anisotropy are generically large relative to the isotropic case. Observe also that these deviations grow with increasing NS mass, which is because anisotropy vanishes in the non-relativists (or low-compactness) limit for the H model. Finally, observe that all panels show similar behavior, namely, each observable decreases as one increases $\lambda_\HH$ for a fixed NS mass. The maximum NS mass also decreases as one increases $\lambda_\HH$, as shown in~\cite{Horvat:2010xf,Doneva:2012rd,Silva:2014fca}. 

\begin{figure}[htb]
\begin{center}
\includegraphics[width=8.5cm,clip=true]{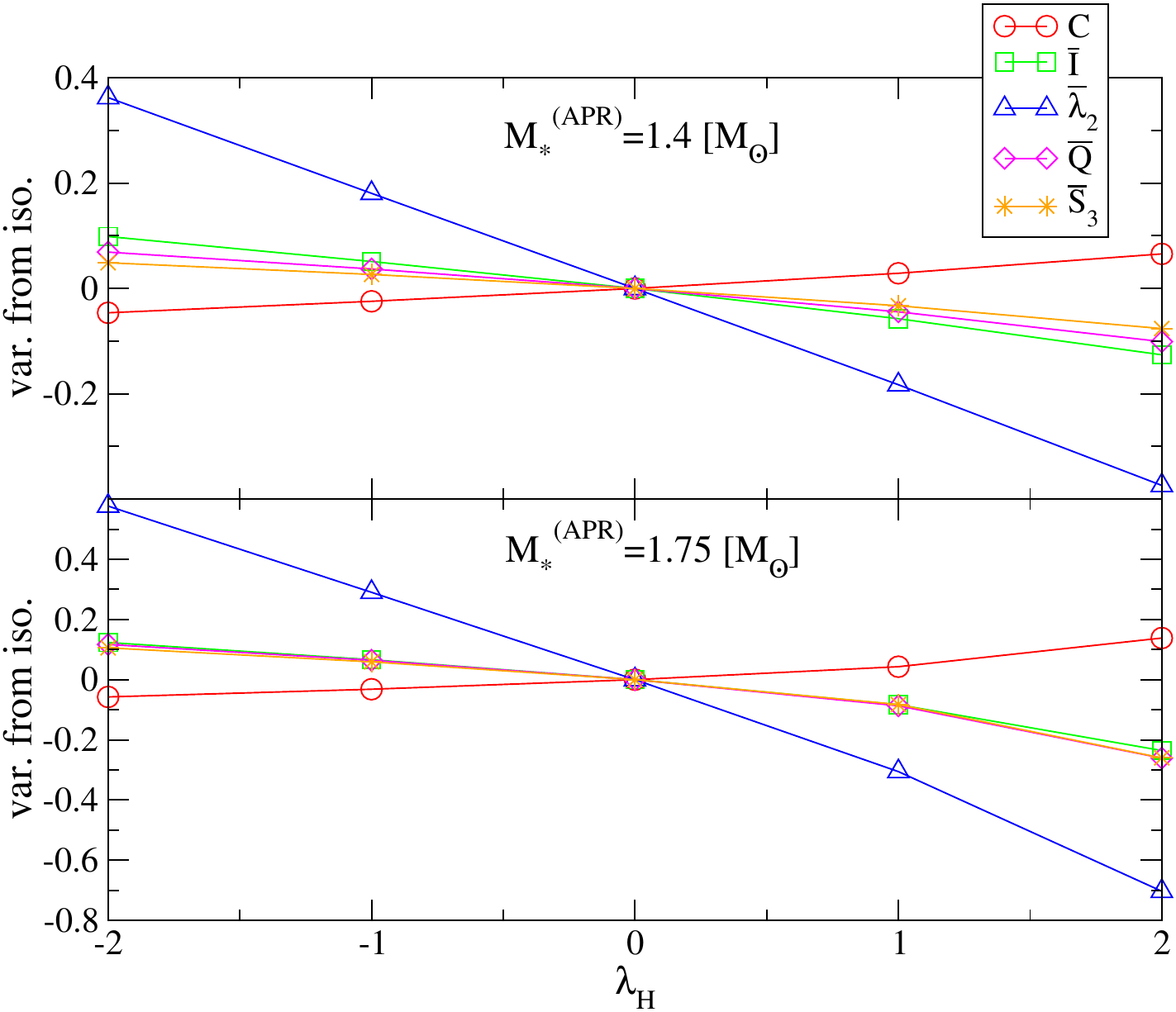}  
\caption{\label{fig:diff-fixed-mass} (Color online) Fractional relative difference of each dimensionless quantity from the isotropic one as a function of $\lambda_\HH$ with $M_* = 1.4M_\odot$ (top) and  $M_* = 1.75M_\odot$ (bottom) for an APR EoS. For reference, we also include the change in stellar compactness, $C$, due to the effect of anisotropy in the stellar radius for a fixed mass. Observe that the effect of anisotropy is almost linear in $\lambda_\HH$.}
\end{center}
\end{figure}
This figure, however, does not allow us to easily quantify the physical effect of anisotropy on compact stars, so let us investigate this more carefully. Figure~\ref{fig:diff-fixed-mass} shows the fractional relative difference between anisotropic and isotropic NSs when calculating $\bar I$, $\bar \lambda_2$, $\bar Q$ and $\bar S_3$, as a function of $\lambda_\HH$, for fixed $M_* = 1.4 M_\odot$ and  $M_* = 1.75 M_\odot$ and an APR EoS.  Observe that the fractional difference scales almost linearly with $\lambda_\HH$. Observe also that the magnitude of the effect of anisotropy for a given $\lambda_\HH$ depends strongly on which quantity one studies. For example, the fractional relative difference in the compactness $C$ and the tidal deformability $\bar \lambda_2$ at $\lambda_\HH = 2$ and $M_* = 1.4 M_\odot$ is 6.6\% and 37\% respectively. 

\begin{figure*}[htb]
\begin{center}
\includegraphics[width=8.5cm,clip=true]{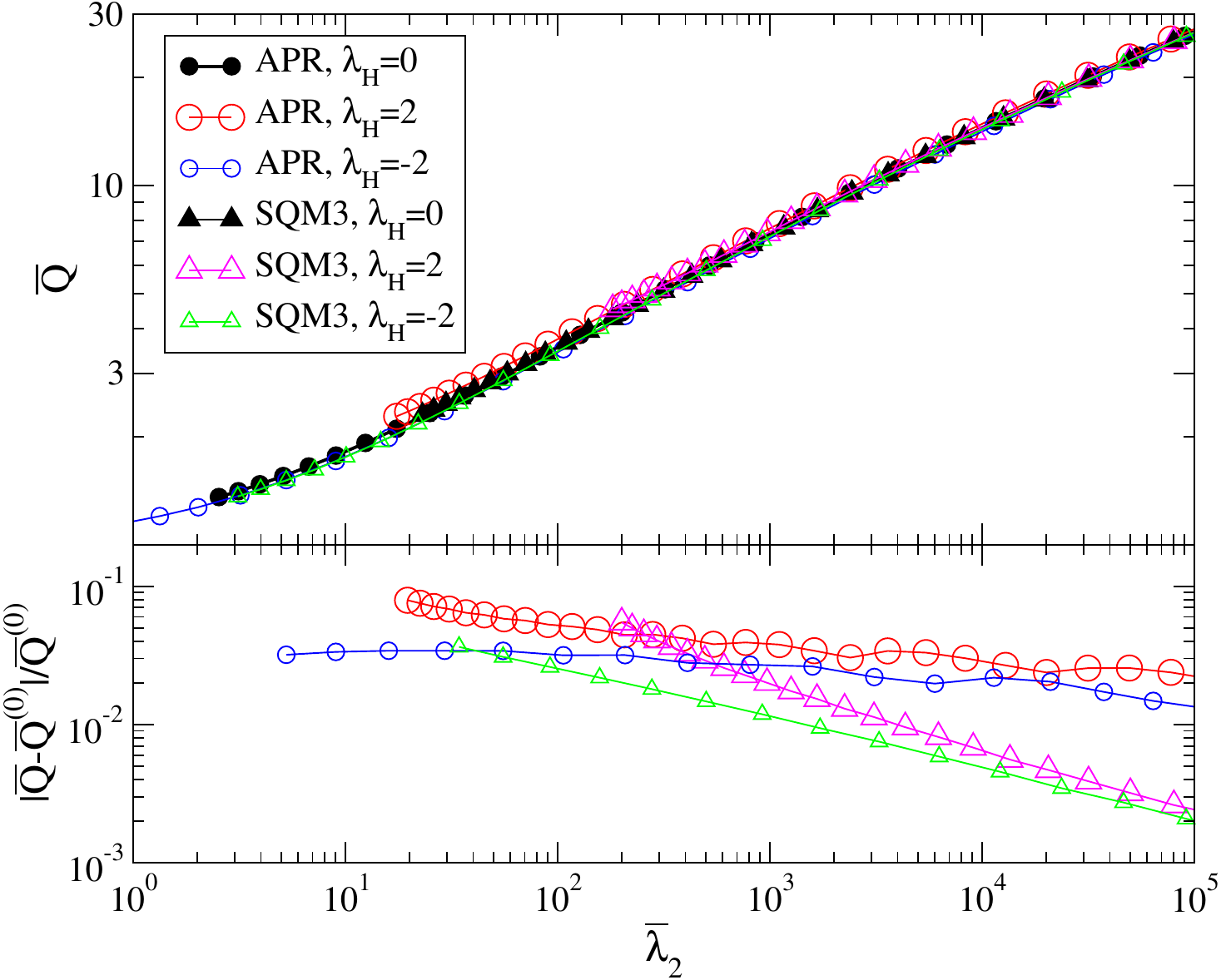}  
\includegraphics[width=8.5cm,clip=true]{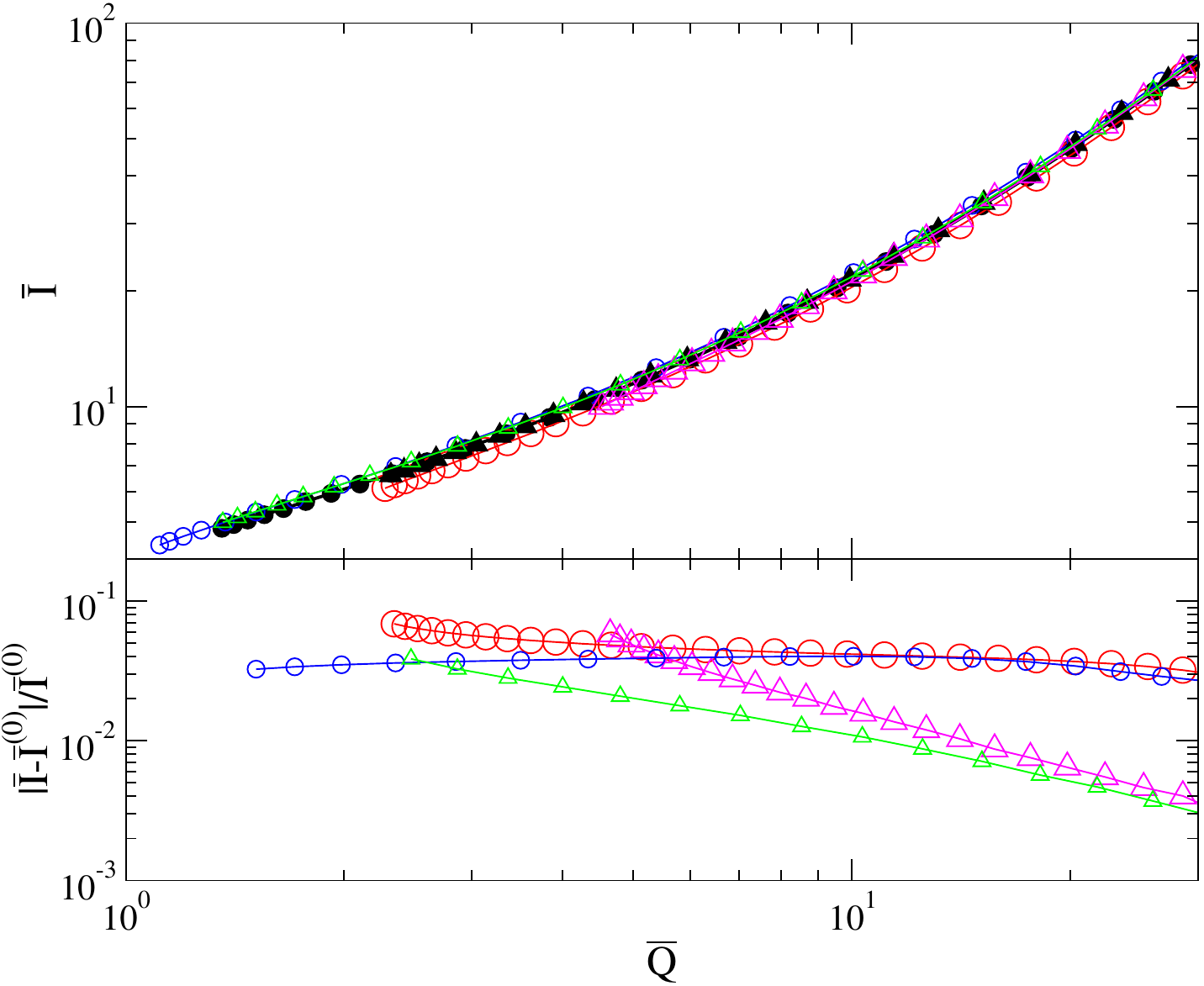}  
\caption{\label{fig:univ-APR-SQM3} (Color online) Same as Fig.~\ref{fig:univ-APR-SQM3-intro} but for the Q-Love (left) and I-Q (right) relations. 
}
\end{center}
\end{figure*}
We are now ready to look at how a non-vanishing $\lambda_\HH$ affects the universal relations. This is shown in the top panels of Figs.~\ref{fig:univ-APR-SQM3-intro} and~\ref{fig:univ-APR-SQM3} for an APR and SQM3 EoS. The bottom panels show the fractional relative difference from the isotropic relation. Observe that the relations are affected at most below the $\sim 10\%$ level when one includes maximal anisotropy. These relations are thus much more EoS-insensitive than Fig.~\ref{fig:M-dep}. The variation induced by anisotropy that scales with mass effectively cancels when one plots one dimensionless observable against another.

Let us investigate the variation in each relation in more detail. First, observe that the variation decreases as one increases $\bar \lambda_2$ or $\bar Q$, which corresponds to decreasing the stellar mass or compactness. This, again, is because anisotropy vanishes in the non-relativistic limit for the H model. Second, observe that all relations have an anisotropy variation that is roughly a factor of 2--4 larger at their maximum relative to that found for isotropic stars~\cite{I-Love-Q-Science,I-Love-Q-PRD,Yagi:2014bxa}. Such an added variability may affect some of the applications of the relations in future observations, as we discuss in detail in Sec.~\ref{sec:applications}.

\begin{figure*}[htb]
\begin{center}
\includegraphics[width=8.5cm,clip=true]{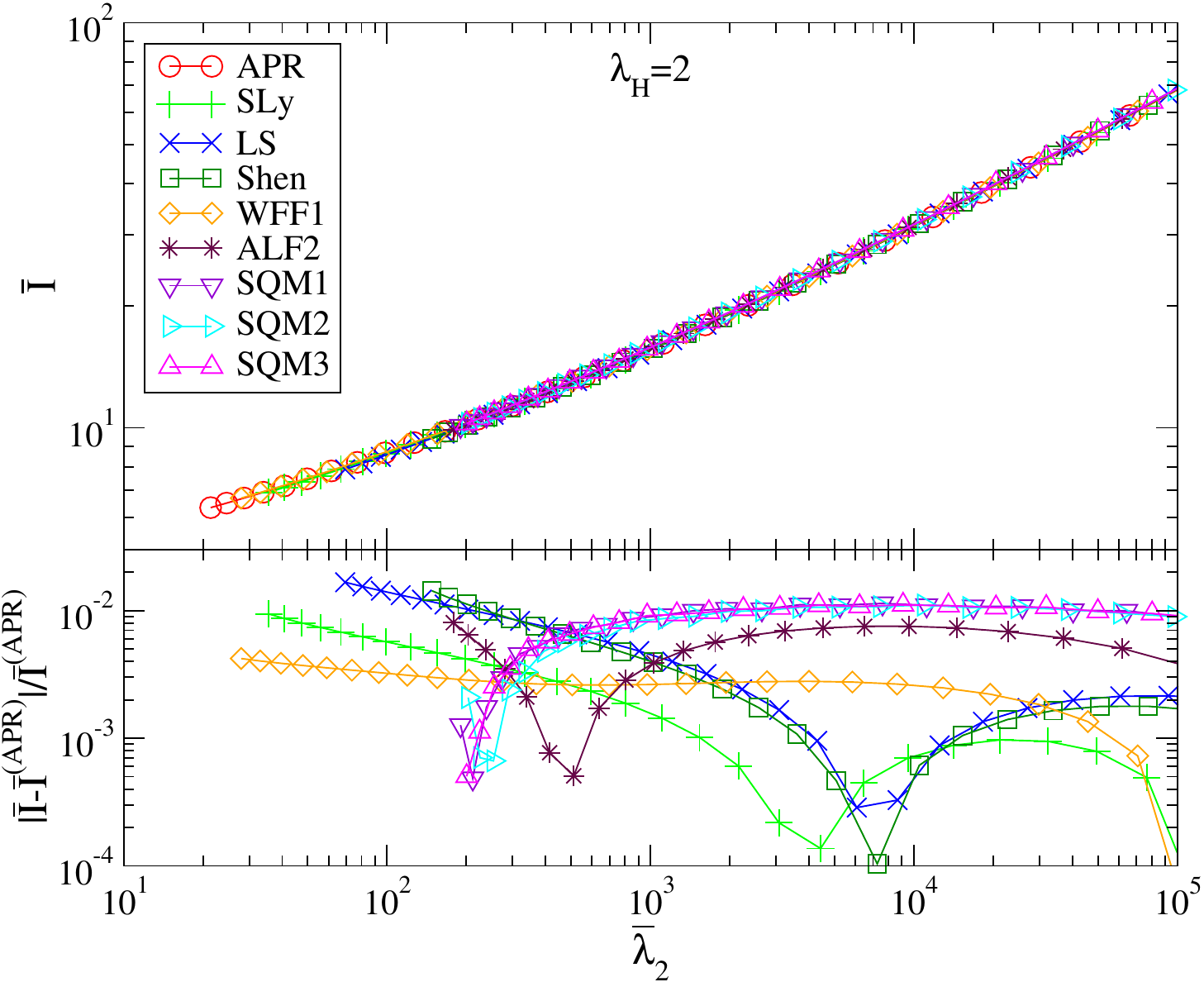}  
\includegraphics[width=8.5cm,clip=true]{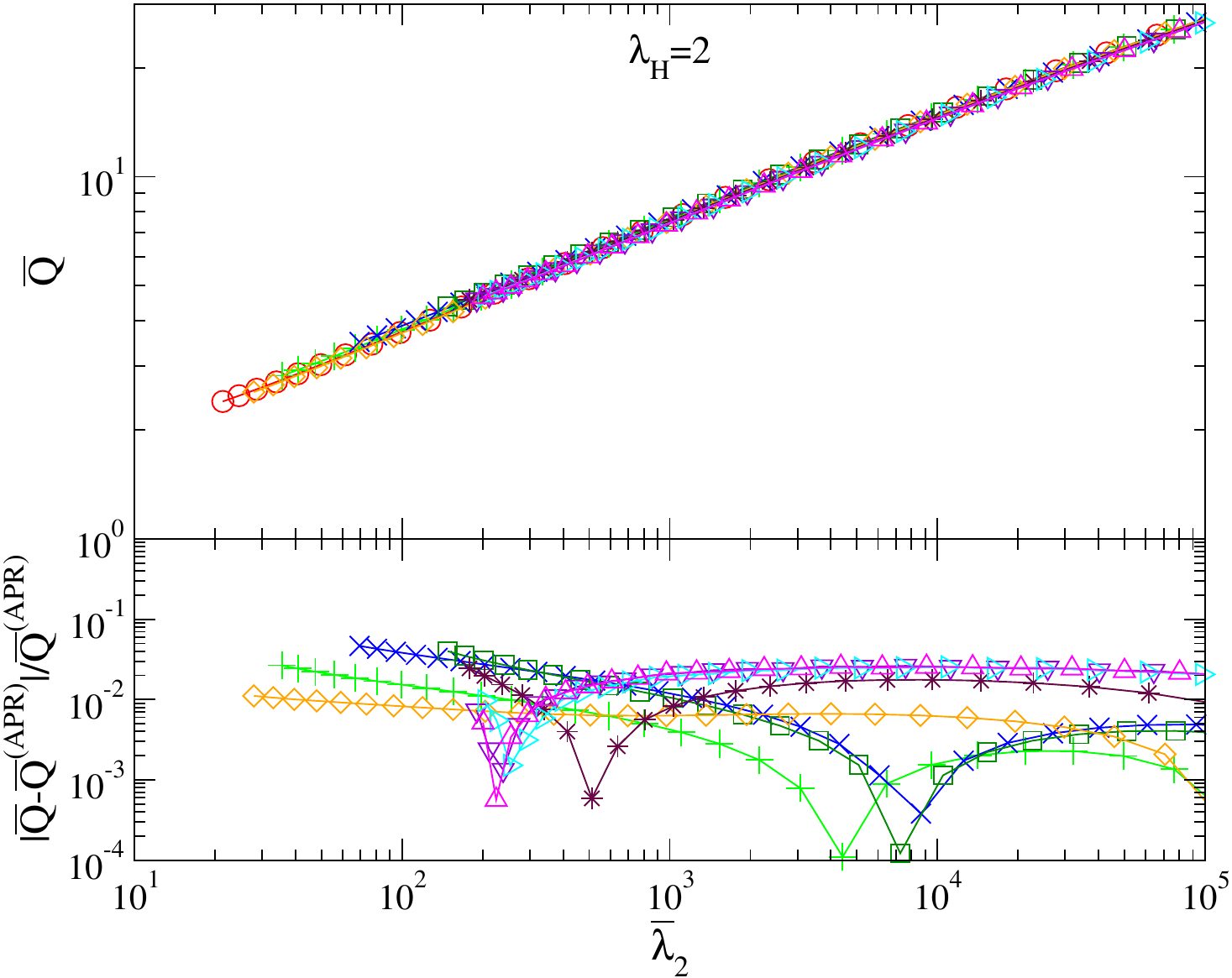}  
\includegraphics[width=8.5cm,clip=true]{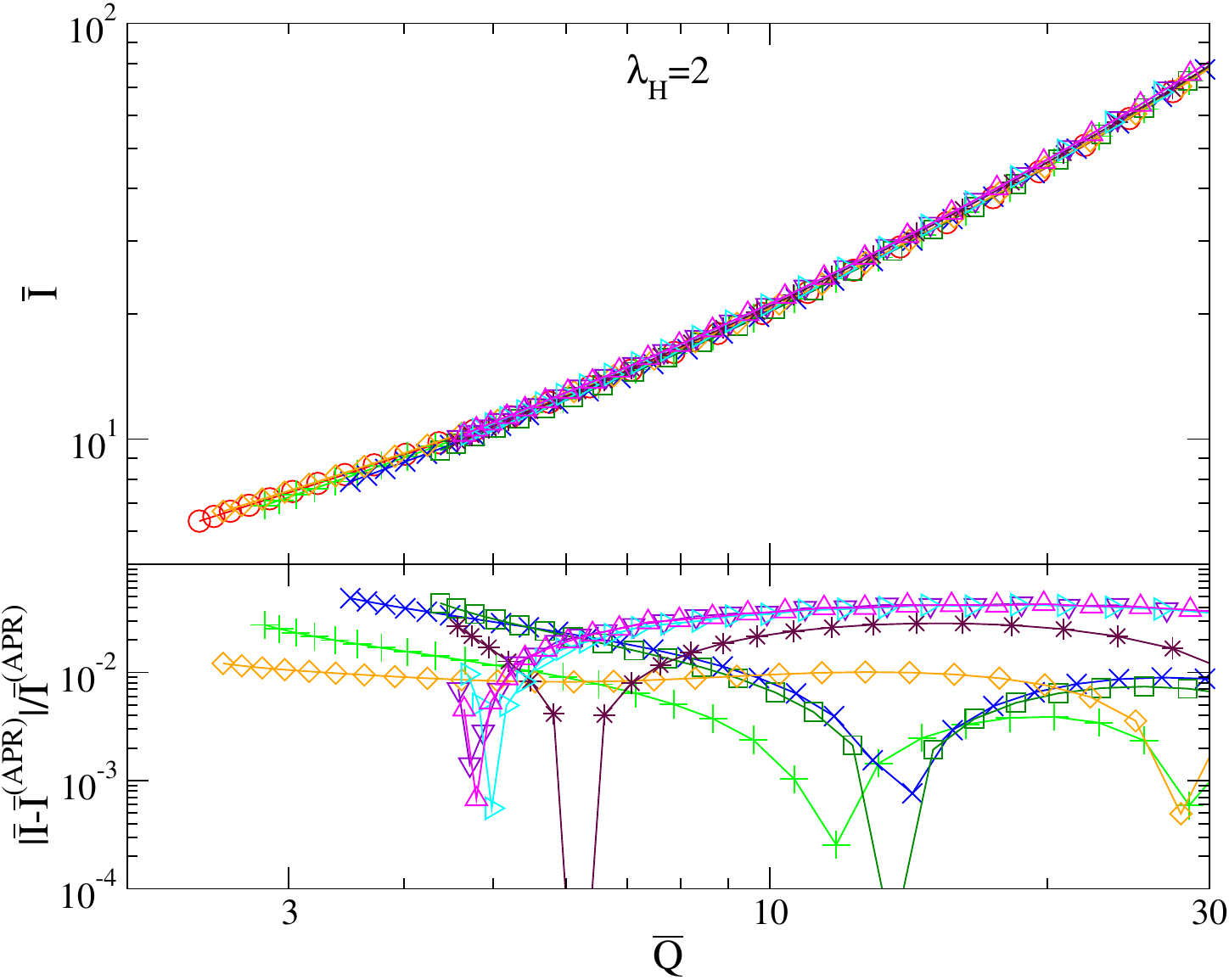}  
\includegraphics[width=8.5cm,clip=true]{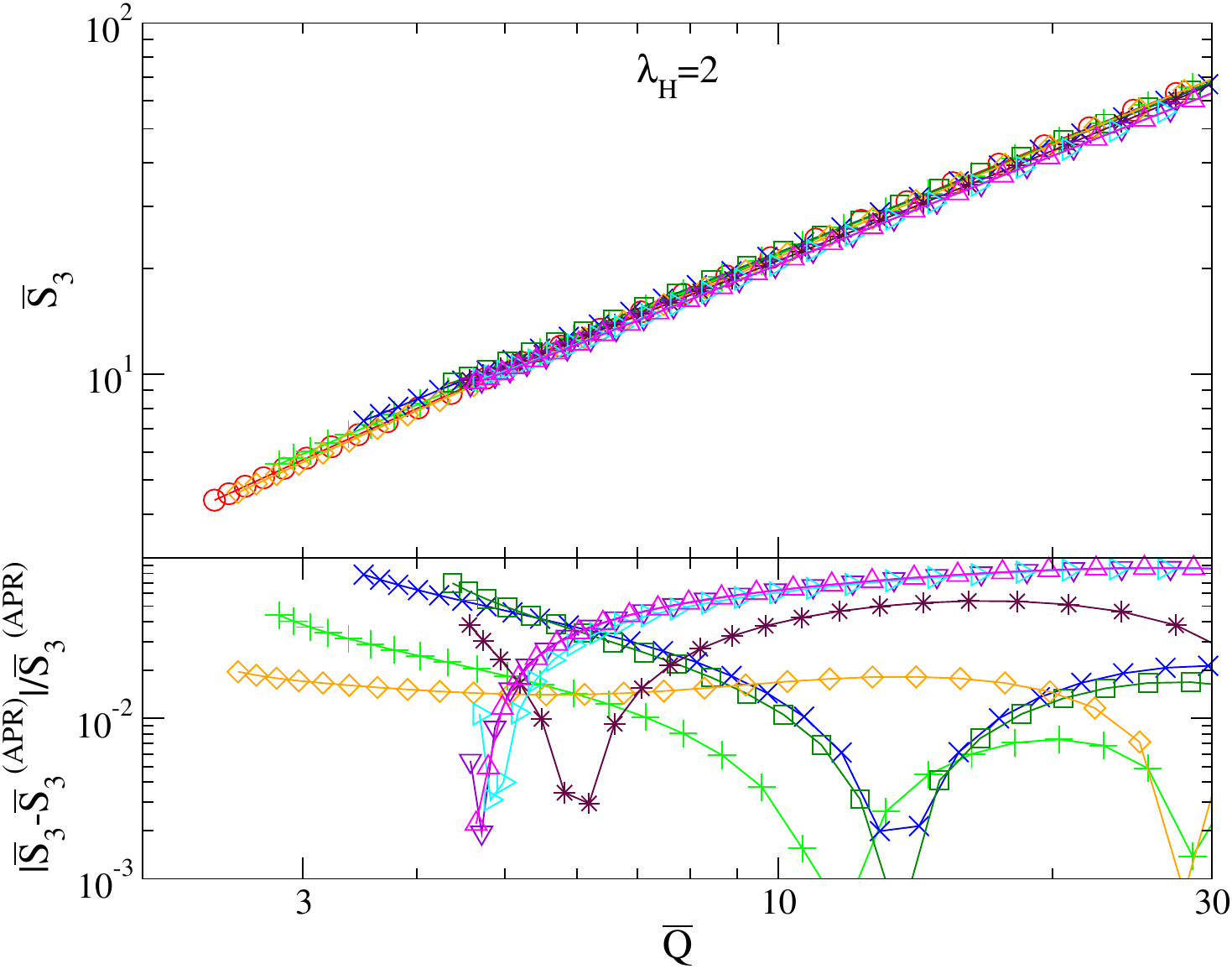}  
\caption{\label{fig:univ-lambda2-2} (Color online) (Top panels) I-Love (top left), Q-Love (top right), I-Q (bottom left) and S$_3$-Q relations for $\lambda_{\HH}=2$ with various EoSs. (Bottom panels) Fractional difference of each relation from that with an APR EoS. Observe that such a EoS variation is comparable to the anisotropy variation shown in Figs.~\ref{fig:univ-APR-SQM3-intro} and~\ref{fig:univ-APR-SQM3}.
}
\end{center}
\end{figure*}
\begin{figure}[htb]
\begin{center}
\includegraphics[width=8.5cm,clip=true]{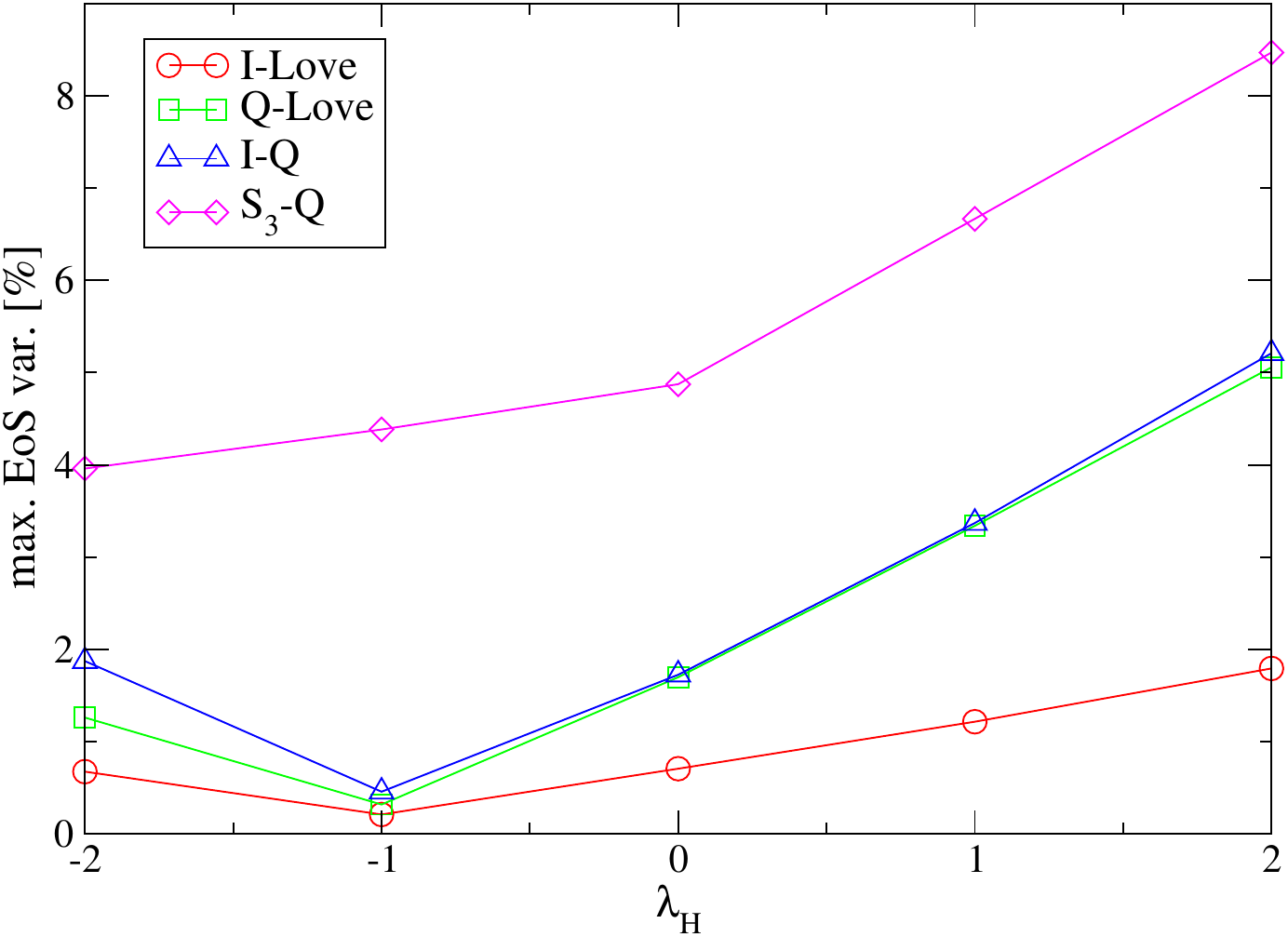}  
\caption{\label{fig:max-diff} (Color online) Absolute value of the maximum EoS variation in the approximate universal relations relative to the APR case as a function of $\lambda_\HH$. Observe that the variation increases almost linearly as one increases anisotropy with positive $\lambda_\HH$.}
\end{center}
\end{figure}

How does anisotropy affect the EoS-variation of the universal relations for fixed anisotropy? The top panels of Fig.~\ref{fig:univ-lambda2-2} present the relations for a fixed $\lambda_\HH=2$ and different EoSs, while the bottom panels show the fractional difference of each relation relative to the APR one. Interestingly, the EoS variation for fixed anisotropy is comparable to the anisotropy variation for fixed EoS, as shown in Figs.~\ref{fig:univ-APR-SQM3-intro} and~\ref{fig:univ-APR-SQM3}. Namely, anisotropy makes the EoS variation slightly larger in some of the relations, but the variation is kept to $\sim 10\%$ at most for the most anisotropic models. In order to see how the EoS variation scales with $\lambda_\HH$, Fig.~\ref{fig:max-diff} presents the absolute maximum fractional difference between the approximate universal relations for each EoS relative to the APR case as a function of $\lambda_\HH.$ Notice that the values at $\lambda_{\HH} = 2$ correspond to those listed on the second row of Table~\ref{table-summary}. Observe that the difference scales almost linearly for positive $\lambda_\HH$ and that the fractional difference becomes largest at $\lambda_\HH = 2$.

\begin{figure*}[htb]
\begin{center}
\includegraphics[width=8.5cm,clip=true]{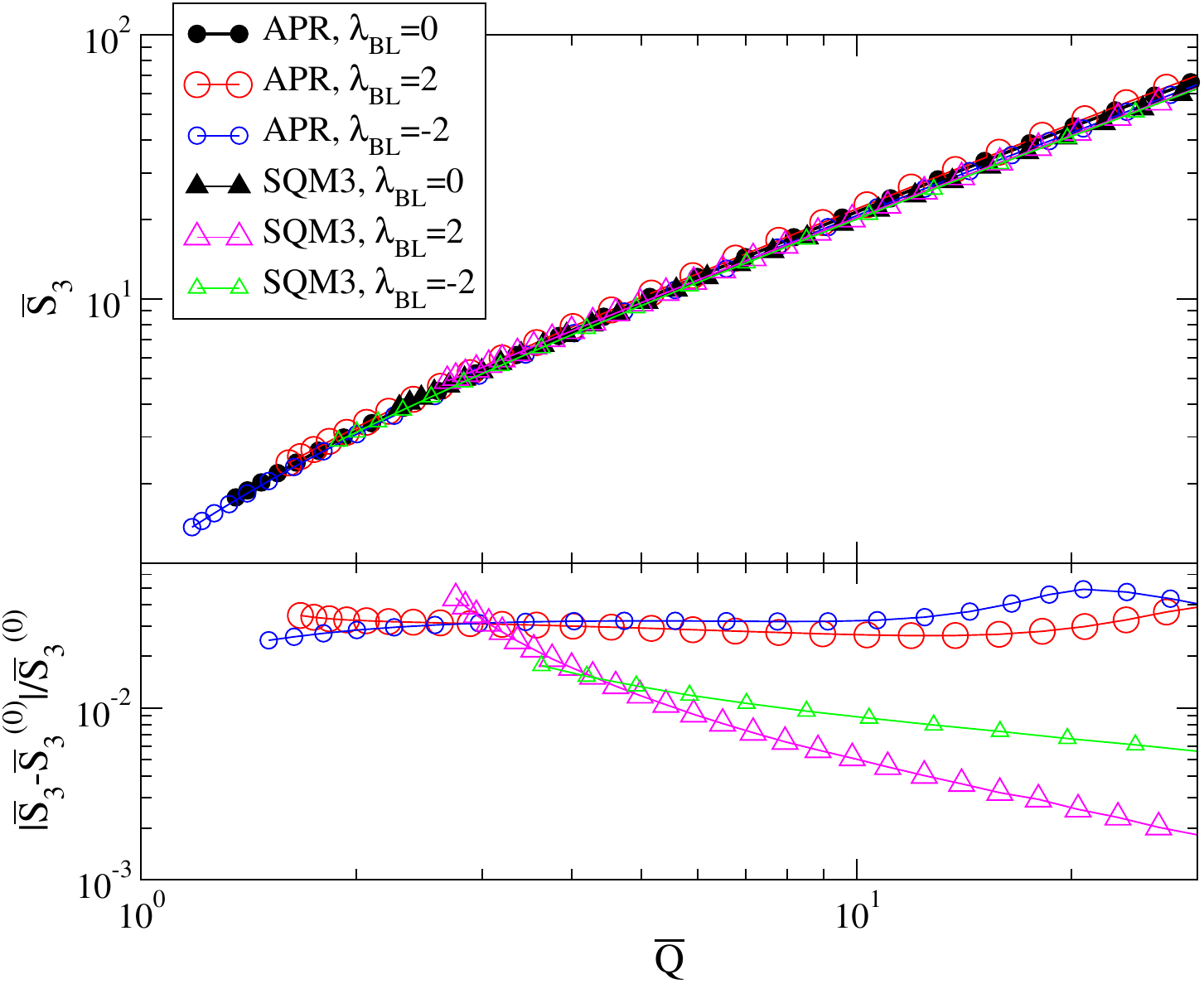}  
\includegraphics[width=8.5cm,clip=true]{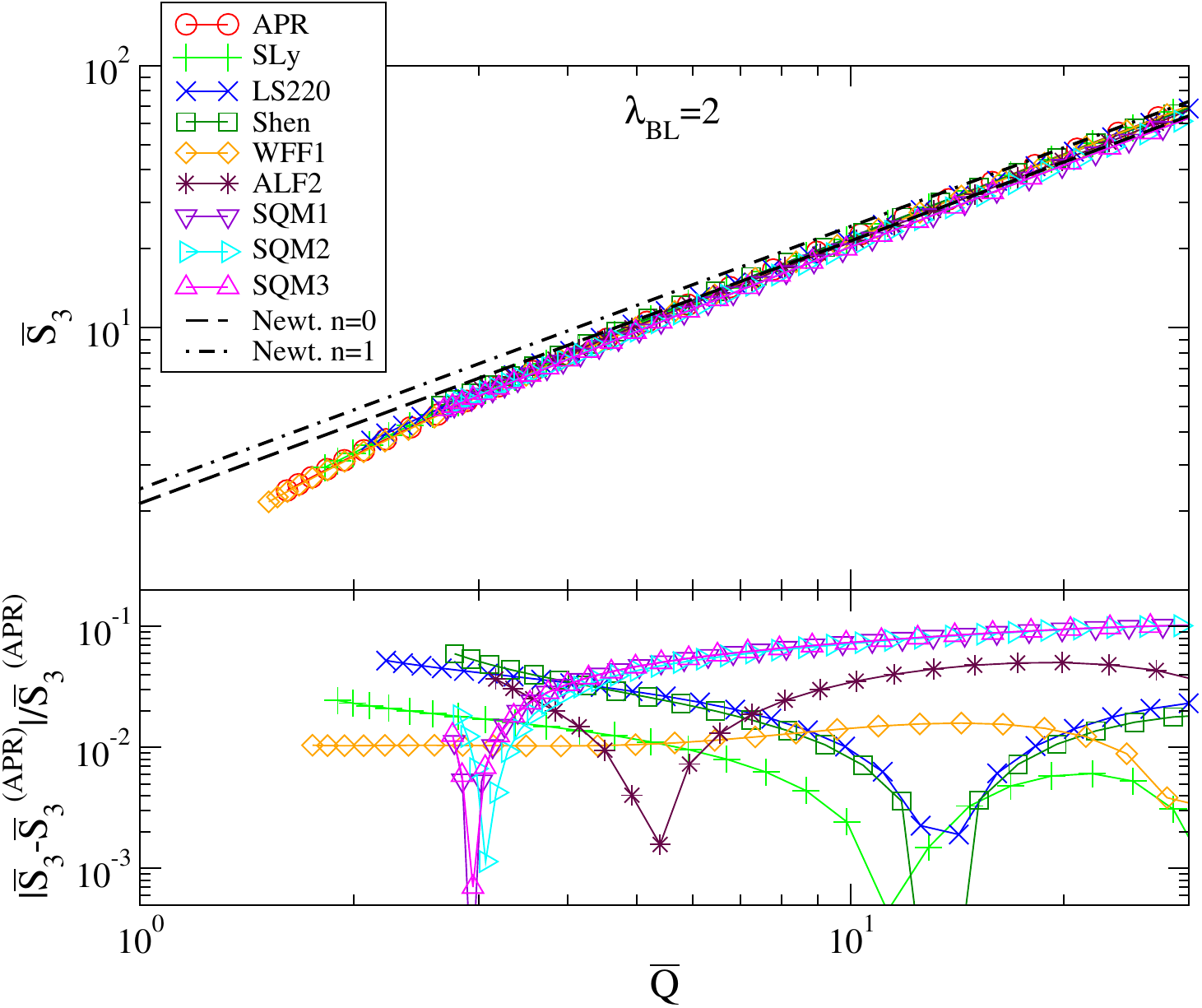}  
\caption{\label{fig:univ-lambda2-BL} (Color online) (Top) S$_3$-Q relations in the BL model for an APR and SQM3 EoSs with various anisotropy parameters $\lambda_\BL$ (left) and for various EoSs with $\lambda_\BL = 2$ (right).  We also present analytic Newtonian relations for the $n=0$ (dashed) and $n=1$ (dotted-dashed) polytropes with $\lambda_\BL=2$ in the top left panel. (Bottom) Fractional difference of each relation from the isotropic case (left) and from that of an APR EoS (right). Observe that the anisotropy and EoS variations are $\sim 5\%$ and $\sim 10\%$ respectively.
}
\end{center}
\end{figure*}
Is the effect of anisotropy in the H model special in some way, or it comparable to other models? To address this question, Fig.~\ref{fig:univ-lambda2-BL} presents the $S_3$-Q relation in the BL model for an APR and SQM3 EoS with various $\lambda_\BL$, while the top right panel shows the relation for various EoSs with fixed $\lambda_\BL = 2$. Observe that the anisotropy and EoS variations are $\sim 5\%$ and $\sim 10\%$ respectively. Indeed, this is very similar to what we found in the H model\footnote{The effect of anisotropy and EoS variation on the I-Love, Q-Love and I-Q relations of NSs is similar in the BL and H models, but this is not the case for QSs. For the latter, such relations show large deviations from the isotropic case, especially in the non-relativistic limit, because of the pathologies of the BL model in this limit.}. 

\section{Physical Interpretation of Universal Relations for Anisotropic Stars}
\label{sec:analytic}

In this section, we provide a physical and a mathematical interpretation of the approximate universality when considering anisotropic stars. We first carry out semi-analytic investigations on the BL model by extending the three-hair relations for isotropic Newtonian polytropes~\cite{Stein:2014wpa} to anisotropic ones in terms of solutions to the modified Lane-Emden (LE) function. We also follow~\cite{Chatziioannou:2014tha} and consider perturbations of the approximate universal relation about an $n=0$ polytropic EoS and obtain completely analytic three-hair relations for anisotropic Newtonian polytropes. These calculations are done within the elliptical isodensity approximation, which assumes that the stellar eccentricity is constant throughout the surface. Here, the stellar eccentricity is a measure of the difference between the radii at the equator and at the pole. We then confirm the accuracy of such an approximation numerically.

\subsection{Three-hair Relations for Anisotropic, Newtonian Polytopes}

Let us start by deriving the three-hair relations for anisotropic polytropes in the Newtonian limit. In such a limit, we showed that for isotropic stars with polytropic EoSs and within the elliptical isodensity approximation, the stellar multipole moments can be expressed only in terms of the first three: the mass, the spin angular momentum and the quadrupole moment~\cite{Stein:2014wpa}. 
In the isotropic case, such three-hair relations are given by 
\be
\label{eq:three-hair}
\bar M_{2\ell +2} + i \bar S_{2 \ell +1} = \bar B_{n,\ell} \bar M_2^\ell \left( \bar M_2 + i \bar S_1 \right)\,,
\ee
or equivalently,
\be
\label{eq:NS-no-hair}
M_{\ell} + i \left(\frac{q}{a}\right) S_{\ell} = \bar B_{n, \lfloor \frac{\ell-1}{2} \rfloor} \; M_0 \; \left(i q \right)^{\ell}\,,
\ee
where $\lfloor x \rfloor$ stands for the largest integer not exceeding $x$, $a \equiv S_1/M_0$ and $q \equiv - i (M_{2}/M_0)^{1/2}$.  The coefficient $\bar B_{n,\ell}$ is given by
\be
\label{eq:Bbar}
\bar B_{n, \ell} = \frac{3^{\ell +1}}{2 \ell +3} \frac{\mathcal{R}_{n,0}^\ell \;  \mathcal{R}_{n, 2\ell +2}}{\mathcal{R}_{n,2}^{\ell +1}}\,,
\ee
where $\vartheta_\LE (\bar{\xi})$ is the LE solution in terms of the LE coordinate $\bar{\xi}$ and $\bar{\xi}_1$ corresponds to the position of the stellar surface. The integral $\mathcal{R}_{n,\ell}$ is given by
\be
\mathcal{R}_{n,\ell} = \int_0^{\bar{\xi}_1} \vartheta_\LE(\bar{\xi})^n \bar{\xi}^{\ell +2} \; d\bar{\xi}\,.
\ee
The $S_3$--$Q$ relation, in particular, is given by $\bar S_3 = \bar B_{n, 1} \bar M_2$ from Eq.~\eqref{eq:three-hair}. All of the EoS dependence is encoded in $\bar B_{n,\ell}$, but we showed in~\cite{Stein:2014wpa} that such dependence is very weak.

\begin{figure*}[htb]
\begin{center}
\includegraphics[width=8.5cm,clip=true]{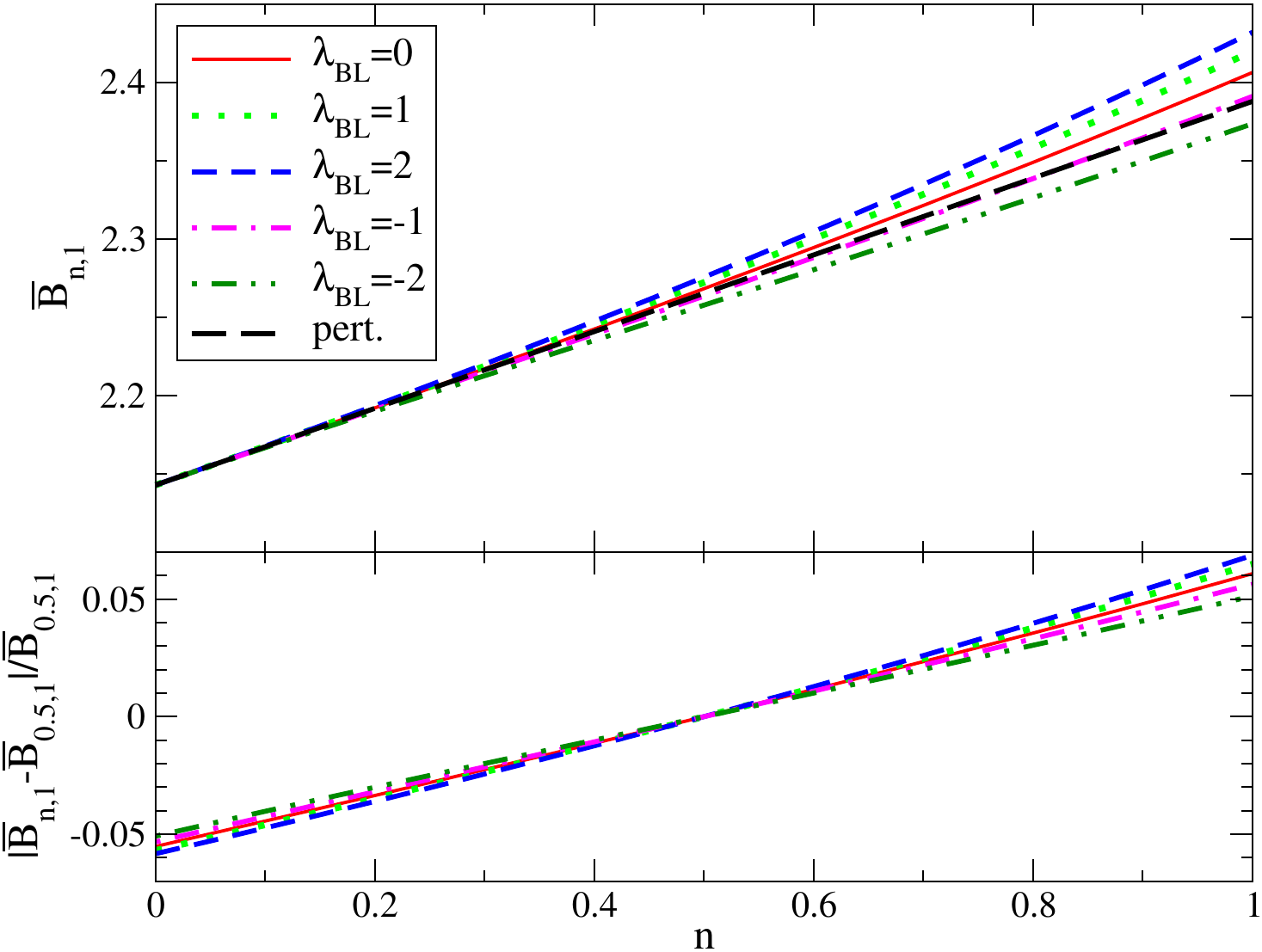}  
\includegraphics[width=8.5cm,clip=true]{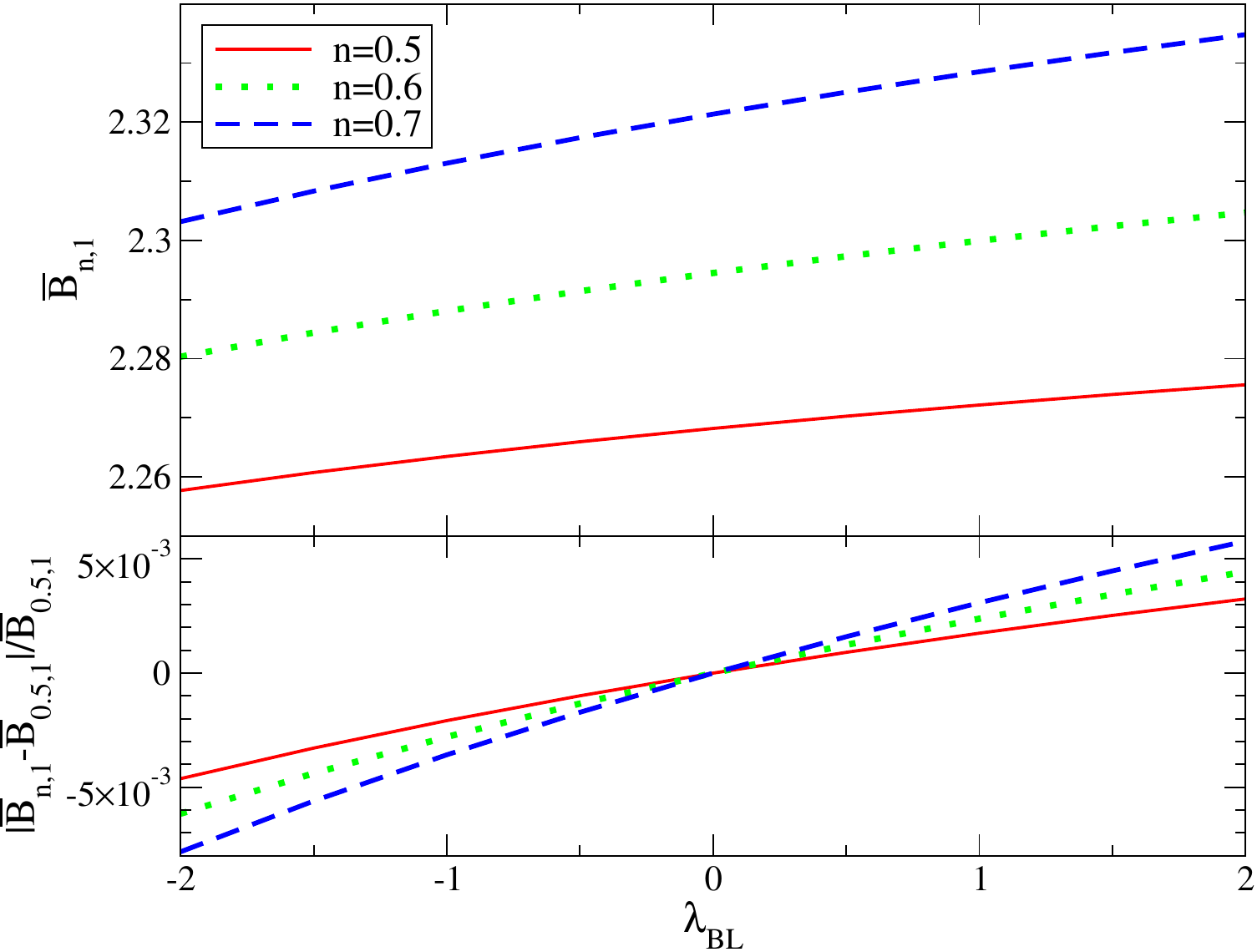}  
\caption{\label{fig:Bbar} (Color online) (Top) $\bar B_{n,1}$ in the S$_3$-Q relation  for Newtonian polytropes with various anisotropy parameters in the BL model as a function of the polytropic index $n$ (left) and with various polytropic indices $n$ as a function of the anisotropy parameter (right). We also show the perturbed $\bar B_{n,1}$ around $n=0$ that is independent of $\lambda_\BL$ in the left panel. (Bottom) Fractional difference of each relation from that for an $n=0.5$ polytrope (left) and  for a polytrope with $\lambda_\BL=0$ (right). Observe that the EoS-variation and anisotropy-variation are within $\sim 7\%$ and $\sim 0.8\%$ respectively even for a large anisotropy case.
}
\end{center}
\end{figure*}

\subsubsection{Modified Lane-Emden Equation and Solution}

With that introduction at hand, let us now concentrate on anisotropic stars with polytropic EoS. From Eqs.~\eqref{eq:dMdR} and~\eqref{eq:dPdR} in the Newtonian limit with the BL model, one finds 
\be
\label{eq:LE}
\frac{1}{\bar{\xi}^2} \frac{d}{d\bar{\xi}} \left[ \bar{\xi}^2 \left( \frac{d\vartheta_\LE}{d\bar{\xi}}  + \frac{\lambda_\BL}{6 \pi}  \bar{\xi} \vartheta_\LE^n \right) \right]  = - \vartheta_\LE^n\,,
\ee
where we used the transformation 
\be
\rho = \rho_c \vartheta_\LE^n\,, \quad r = \sqrt{\frac{n+1}{4 \pi} K \rho_c^{-1 + 1/n}} \; \bar{\xi}\,.
\ee
Equation~\eqref{eq:LE} can also be derived by using the LE equation for a generic anisotropic polytrope, as in~\cite{Herrera:2013dfa}. (See also~\cite{Shojai:2015hsa} for a recent related work on solving the LE equation for Newtonian polytropes with different anisotropy models.)
Obviously, Eq.~\eqref{eq:LE} reduces to the standard LE equation when $\lambda_\BL = 0$. As in the isotropic case, Eq.~\eqref{eq:LE} admits analytic solutions for the $n=0$ and $n=1$ polytropes:
\begin{align}
\vartheta_\LE^{(n=0)} (\bar{\xi}) &= 1 - \frac{\bar{\xi}^2}{6} - \frac{\lambda_\BL \bar{\xi}^2}{12 \pi}\,, \\
\vartheta_\LE^{(n=1)} (\bar{\xi}) &= {}_1 F_1 \left( \frac{3}{2} + \frac{3 \pi}{\lambda_\BL} ; \frac{3}{2} ; - \frac{\lambda_\BL \bar{\xi}^2}{12 \pi} \right)\,, 
\end{align}
where ${}_1 F_1 (a;b;z)$ is the Kummer confluent hypergeometric function.

Let us investigate the three-hair relations for anisotropic, incompressible $(n=0)$ polytropes in more detail. In this case, one finds
\be
\label{eq:R-xi1-n0}
\mathcal{R}_{0,\ell} = \frac{\bar \xi_1^{\ell+3}}{\ell + 3}\,, \quad \bar{\xi}_1^{(n=0)} =\frac{2 \sqrt{3 \pi}}{\sqrt{2\pi + \lambda_\BL}}\,,
\ee
and 
\be
\label{eq:theta1-n0}
\vartheta_\LE^{(n=0)}{}'(\bar \xi_1) = -\frac{\sqrt{2\pi + \lambda_\BL}}{\sqrt{3 \pi}}\,.
\ee
Substituting Eqs.~\eqref{eq:R-xi1-n0} and~\eqref{eq:theta1-n0} into Eq.~\eqref{eq:Bbar}, one finds
\be
\label{eq:Bbar-n0}
\bar B_{0,\ell} =  \frac{3\ 5^{\ell+1}}{4 \ell^2+16 \ell+15}\,.
\ee
Observe that $\bar B_{0,\ell}$ is independent of $\lambda_\BL$, and hence, the three-hair relations for incompressible anisotropic polytropes are the same as those for isotropic ones.

The top panel of Fig.~\ref{fig:Bbar} presents $\bar B_{n,1}$ in the $S_3$--$Q$ relation as a function of $n$ for fixed $\lambda_\BL$ (left) and as a function of $\lambda_{\BL}$ for fixed $n$ (right). These figures are generated by solving the modified LE equation in Eq.~\eqref{eq:LE} numerically with the boundary conditions $\vartheta_\LE (0) = 1$ and $\vartheta_\LE' (0) = 0$. Observe that $\bar B_{n,1}$ is almost independent of $\lambda_\BL$ at $n \sim 0$, consistent with our finding of $\bar B_{0,\ell}$ described above. Observe also that the EoS variation is smaller (larger) as one decreases (increases) $\lambda_\BL$. Such a behavior is consistent with Fig.~\ref{fig:max-diff}, where the latter also includes relativistic effects.

The bottom panel of Fig.~\ref{fig:Bbar}  shows the fraction difference from $\bar B_{\langle n \rangle, 1}$, where $\langle n \rangle$ is the averaged polytropic index considered, which we take as $\langle n \rangle = 0.5$. Observe that the EoS dependence is $\sim 7\%$ even when $\lambda_\BL = 2$, whereas the EoS variation is $\sim 6\%$ for isotropic polytropes. Therefore, this calculation explains why the EoS variation in the S$_3$-Q relation for anisotropic stars is comparable to that for isotropic ones.
 Such Newtonian $S_3$--$Q$ relations for the $n=0$ and $n=1$ polytropes with $\lambda_\BL = 2$ are shown in Fig.~\ref{fig:univ-lambda2-BL} as dashed and dotted-dashed lines respectively. Observe how numerical results approach these Newtonian lines as one decreases the stellar compactness.

\begin{figure*}[htb]
\begin{center}
\includegraphics[width=8.5cm,clip=true]{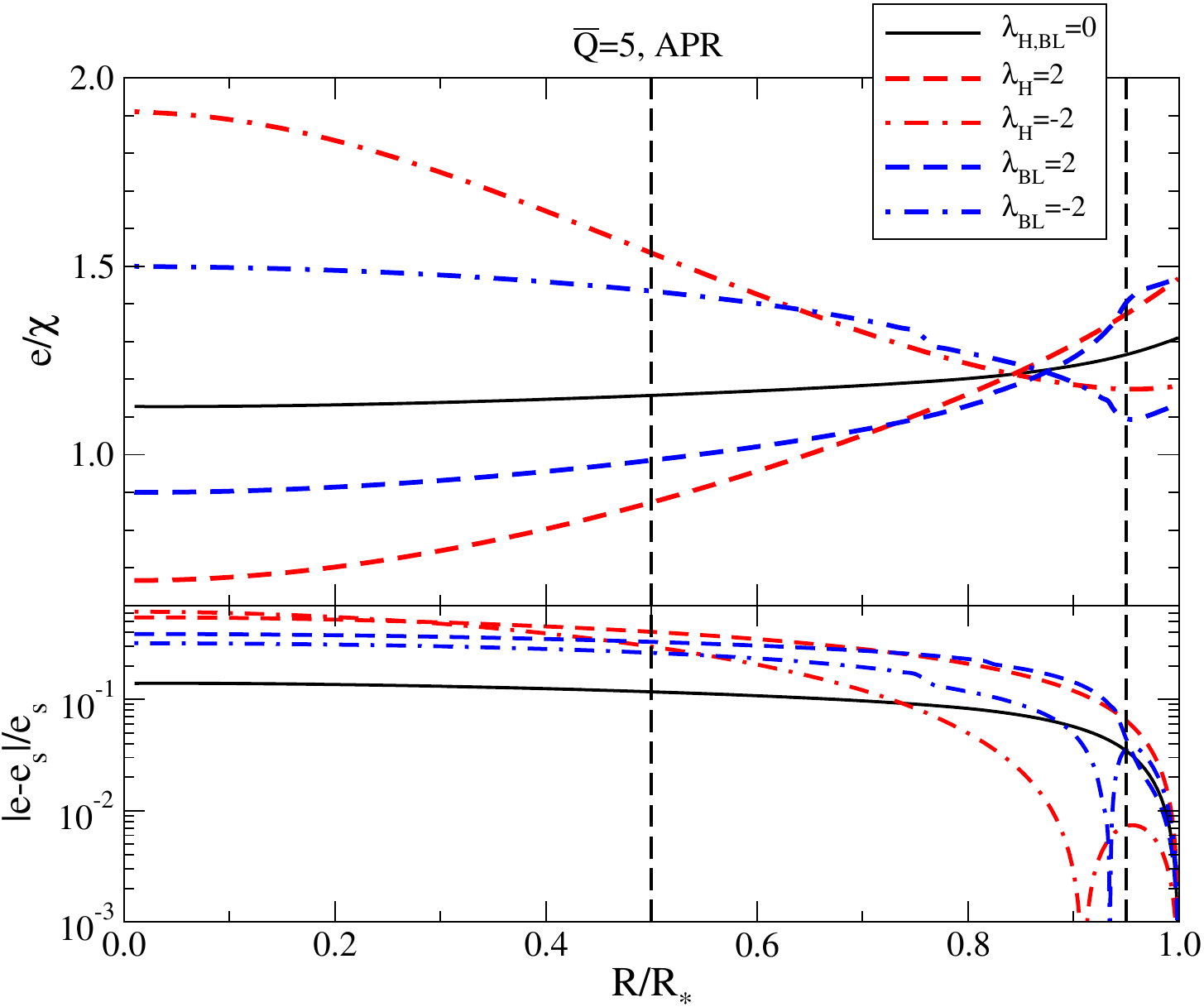}  
\includegraphics[width=8.5cm,clip=true]{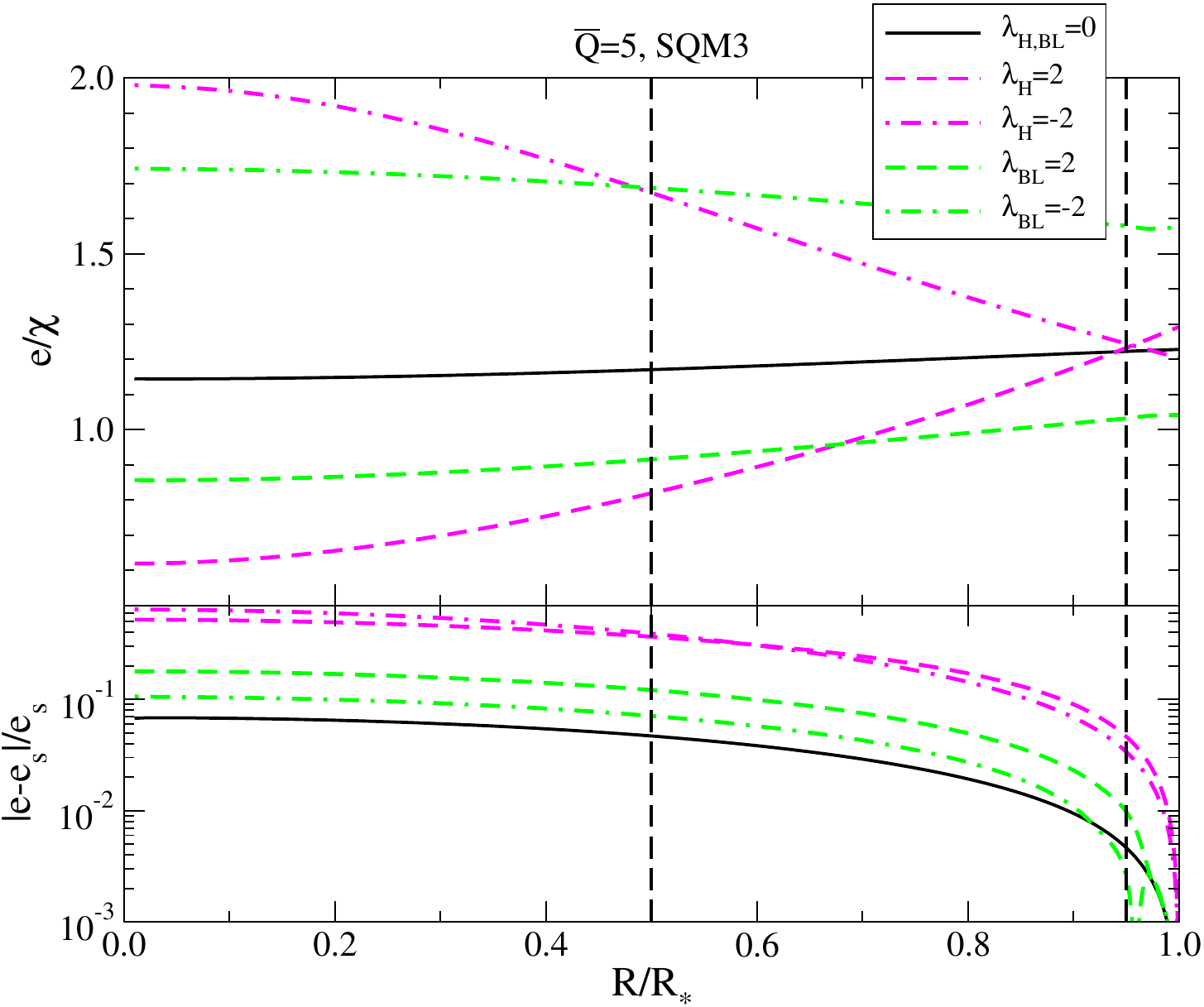}  
\caption{\label{fig:ecc} (Color online) (Top panels) Eccentricity profile of NSs with an APR EoS (left) and QSs with SQM3 EoS (right) for various anisotropy parameters with $\bar Q =5$. (Bottom panels) Fractional difference of each profile from the surface value $e_s$. The region within dashed lines correspond to the part that contributes the most to the EoS-universality in the relations among $\bar I$ and $\bar Q$~\cite{Yagi:2014qua}.
}
\end{center}
\end{figure*}

\subsubsection{Perturbation about Incompressible Stars}

Let us now try to construct analytically three-hair relations for anisotropic stars for other polytropic EoSs. To do so,  we now follow~\cite{Chatziioannou:2014tha} and solve Eq.~\eqref{eq:LE} analytically by considering a perturbation around a polytrope with a fiducial polytropic index $\tilde n$. We start by decomposing $n$ and $\vartheta_\LE$ as
\be
n = \tilde n + \epsilon_n\,, \quad \vartheta_\LE = \tilde{\vartheta}_\LE + \epsilon_n \delta \vartheta_\LE + \mathcal{O}\left( \epsilon_n^2 \right)\,,
\ee
where $\epsilon_n$ is a perturbation to the background polytropic index $\tilde n$, while $\delta \vartheta_\LE$ is a perturbation to the background solution to the LE equation $\tilde{\vartheta}_\LE$. We now set the background $\tilde n =0$ and consider perturbations to the incompressible, anisotropic Newtonian star. The linear order perturbation to the LE equation with $\tilde n = 0$ is given by 
\be
\frac{1}{\bar{\xi}^2} \frac{d}{d\bar{\xi}} \left( \bar{\xi}^2 \frac{d \delta \vartheta_\LE}{d\bar{\xi}} \right) + \ln \tilde \vartheta_\LE + \frac{\lambda_\BL}{6\pi} \left( 3 \ln \tilde \vartheta_\LE + \frac{\bar{\xi} \tilde \vartheta_\LE'}{\tilde \vartheta_\LE} \right)=0\,.
\ee
The solution to the perturbed LE equation above under the boundary condition $\delta \vartheta_\LE (0) = 0 = \delta \vartheta_\LE'(0)$ is given by
%
\begin{align}
\delta \vartheta_\LE (\bar{\xi}) &= \frac{1}{36 \pi  (2 \pi + \lambda_\BL )^{3/2} \bar{\xi} } \left\{ 576 \sqrt{3} \pi ^{5/2} 
\right. 
\nn \\
& \left.
\tanh ^{-1}\left(\frac{\sqrt{2 \pi + \lambda_\BL }  }{2 \sqrt{3 \pi }} \bar{\xi} \right) - \sqrt{2 \pi + \lambda_\BL } \bar{\xi}  \left[ 288 \pi ^2  
\right. \right. \nn \\ 
&  \left.\left. 
- (2 \pi + \lambda_\BL) (10 \pi + 3 \lambda_\BL ) \bar{\xi} ^2  - 3 \left[ 12 \pi  (6 \pi + \lambda_\BL ) 
\right. \right. \right. \nn \\ 
&  \left.\left. \left.
- (2 \pi + \lambda_\BL )^2 \bar{\xi} ^2 \right] \ln \left(1-\frac{2 \pi + \lambda_\BL}{12 \pi } \bar{\xi} ^2 \right) \right]  \right\}\,,
\end{align}
%
while stellar surface is given by 
\be
\bar{\xi}_1 = \bar{\xi}_1^{(n=0)} \left(  1 +  \frac{ 2 \pi  (12 \ln 2 -7) + 3 \lambda_\BL}{6 (2 \pi + \lambda_\BL)}  \epsilon_n  \right)+ \mathcal{O}\left( \epsilon_n^2 \right)\,,
\ee
where $ \bar{\xi}_1^{(n=0)}$ is given by Eq.~\eqref{eq:R-xi1-n0}.

With these perturbed solutions at hand, one can now calculate the perturbed $\bar B_{n,\ell}$ as~\cite{diff-rot}
\begin{align}
\label{eq:Bbar-pert}
\bar B_{n,\ell} &= \bar B_{0,\ell} \left\{ 1+ \left[ \frac{46}{15} + \frac{2 \ell}{5} - 2 \ln 2 - H \left( \frac{5}{2} + \ell \right) \right] \epsilon_n \right\}  \nn \\
&+ \mathcal{O}\left( \epsilon_n^2 \right)\,, 
\end{align} 
where $H(x)$ is the $\ell$th Harmonic number and $\bar B_{0,\ell}$ is given by Eq.~\eqref{eq:Bbar-n0}. Notice that $\bar{B}_{n,\ell}$ does not depend on $\lambda_\BL$, and hence, the relation is the same as that for isotropic stars. This analytic result explains mathematically why $\bar B_{n,\ell}$ is insensitive to $\lambda_\BL$ close to $n \sim 0$. In the top panel of Fig.~\ref{fig:Bbar}, we also show the Newtonian $S_3$--$Q$ relation with $\bar B_{n,1}$ given in Eq.~\eqref{eq:Bbar-pert} as a black dashed line. Of course, this perturbed $\bar{B}_{n,\ell}$ cannot accurately describe the $\lambda_{\BL} = 0$ line that is obtained numerically in the isotropic case, because the former is only an analytic expansion of the latter to linear order in the perturbation. Interestingly, such a relation is most suited to describe the numerical relation up to $n=1$ with $\lambda_\BL = -1$. 

\subsection{Eccentricity Profile of Anisotropic Stars}

Let us now try to understand physically why anisotropy increases the degree of variability of the approximate universal relations. Reference~\cite{Yagi:2014qua} proposed a phenomenological explanation for the appearance of universality: the emergence of an approximate symmetry in the form of self-similarity of isodensity contours. As one considers stars with increasing compactness, the eccentricity of the stellar surface decreases inside the star (within the region inside the star, $r \in (0.5,0.95)R_{*}$, that matters the most for the computation of the multipole moments), leading to self-similar contours of constant density, just like the layers of an onion. If this phenomenological model is correct, one would then also expect that the introduction of pressure anisotropy forces an increased variability in the eccentricity profile inside the star.

In order to confirm the above expectation, let us study the eccentricity profile of anisotropic stars constructed numerically. Following~\cite{Hartle:1968ht}, we calculate the stellar eccentricity $e$ via
\begin{align}
e &= \sqrt{\frac{\mrm{(radius \ at \ equator)}^2}{\mrm{(radius \ at \ pole)}^2}-1} 
= \sqrt{-3\left( k_2 + \frac{\xi_2}{R} \right)}\,,
\end{align}
where recall that $k_{2}$ is a metric function [see Eq.~\eqref{eq:metric-ansatz}], while $\xi_{2}$ is a function in the coordinate transformation [see Eq.~\eqref{eq:coord-transf}]. Of course, the parenthesis inside the square root is always negative provided the star is oblate, i.e.~the radius at the equator is larger than the radius at the pole, so the eccentricity is always a real number. Thus, we expect $e=0$ for a sphere, $e>0$ and real for an oblate spheroid. In particular, $e>1$ for a particularly oblate object.

The top panels of Fig.~\ref{fig:ecc} shows the eccentricity profile of compact stars with an APR (left) and SQM3 (right) EoS for both the H and BL models with various $\lambda_\HBL$ at $\bar Q = 5$. Vertical lines show the region that matters the most to the universality.  Comparing the anisotropic curves to the isotropic ones, one sees that the eccentricity variation is much larger for anisotropic stars. Interestingly, the eccentricity increases (decreases) as one increases $R/R_*$ when $\lambda_\HBL$ is positive (negative).

The bottom panels of Fig.~\ref{fig:ecc} show the fractional difference of the stellar radius at each $R/R_*$ from the surface value. Observe that the eccentricity variation of anisotropic stars can be as large as $\sim 35\%$, compared to $\sim 10\%$ variation for isotropic stars. Such an enhancement in the variation is consistent with the EoS variation in the universal relations for anisotropic stars being larger than those for isotropic ones. On the other hand, a $\sim 35\%$ variation in the eccentricity profile means that the EoS variation derived for the three-hair relations within the elliptical isodensity approximation in the previous subsection may have an error of the same order. Therefore, the EoS variation in the S$_3$-Q relation for the BL model should be $7\% \times 1.35 \sim 9.45 \%$ in the non-relativistic limit, which agrees quite accurately with the numerical calculation shown in the bottom right panel of Fig.~\eqref{fig:univ-lambda2-BL}. These findings give further evidence in favor of the claim in~\cite{Yagi:2014qua} that the origin of the approximate universality is the approximate symmetry of the eccentricity profile of isodensity contours. 
 
\section{Applications of Universal Relations}
\label{sec:applications}

In this section, we address the second question posed in Sec.~\ref{sec:intro}, namely, how the effect of anisotropy on the universal relations affects their application to future observations. We look at applications to GW astrophysics, X-ray nuclear astrophysics and experimental relativity in turn.

\subsubsection{Gravitational Wave Astrophysics}

In~\cite{I-Love-Q-Science,I-Love-Q-PRD}, we showed that the Q-Love relation can be useful for measuring the NS spins in GW observations. The reason for that is the following. The gravitational waves emitted by a NS/NS binary in its inspiral phase of evolution can be well modeled in the post-Newtonian (PN) framework, where one expands in the ratio of the velocity of the binary constituents to the speed of light. The spin angular momenta of the NSs enter the gravitational wave phase first at 1.5PN order through particular projections to the orbital angular momentum vector\footnote{Given a PN expansion of a given quantity, a term in the PN series is said to be of $N$PN.}. Given a binary with spins perpendicular to the orbital plane, the spins enter the phase at 1.5PN through their symmetric combination, and at 2PN order through their antisymmetric combination.  One may thus expect to be able to measure the individual spins of such a system given a gravitational wave detection if one can extract the 1.5PN and 2PN terms in the phase. 

The quadrupole moment of each NS, however, is strongly degenerate with the spin angular momenta. Indeed, the quadrupole moment enters the gravitational wave phase first at 2PN order~\cite{poisson-quadrupole,vasuth-spinspin}. This strong degeneracy makes it impossible to measure the individual spins independently from just the 1.5 and 2PN terms in the phase. One could try to use the 2.5PN term in the phase, but the higher the PN order of a given term, the smaller its effect in the inspiral phase, and thus, terms of such high PN order are difficult to extract.

Enter the Q-Love relation. Through this relation one can rewrite the quadrupole moment in terms of the tidal deformability parameter. The latter first enters the gravitational wave phase at 5PN order, which at first sight seems too small an effect to be measurable. However, the tidal deformability enters the phase multiplied by a large coefficient, due to the large compactness of NSs, which may render it measurable~\cite{flanagan-hinderer-love,read-love,hinderer-lackey-lang-read,lackey,damour-nagar-villain,delpozzo,Read:2013zra,Lackey:2014fwa,2010PhRvL.105z1101B,2013PhRvD..87d4001H,2014MNRAS.437L..46R,2014PhRvL.112t1101B,2015PhRvL.114p1103B}. Thus, if one can extract the tidal deformability from the 5PN term in the phase, and then use this to infer the value of the quadrupole moment in the 2PN term, one can then use the 1.5 and 2PN terms in the phase to extract the individual spins. More precisely, the re-expression of the quadrupole moment in terms of the tidal deformability eliminates one degenerate direction from the template or model submanifold, thus allowing us to identify and explore directions that were previously inaccessible. 

Let us now discuss the applicability of the Q-Love relation to GW observations when anisotropy is present. 
Figures~\ref{fig:univ-APR-SQM3} and~\ref{fig:univ-lambda2-2} show that the anisotropy and EoS variations in the Q-Love relation for anisotropic stars are $\sim 10\%$. Therefore, we conclude that even if anisotropy were present, one could still apply the relation to GW observations as long as the measurement accuracy of the spins is no better than $\sim 10\%$. Such a measurement accuracy is what one expects from GW observations~\cite{I-Love-Q-Science,I-Love-Q-PRD}.

\subsubsection{X-ray Nuclear Astrophysics}

\begin{figure}[htb]
\begin{center}
\includegraphics[width=8.5cm,clip=true]{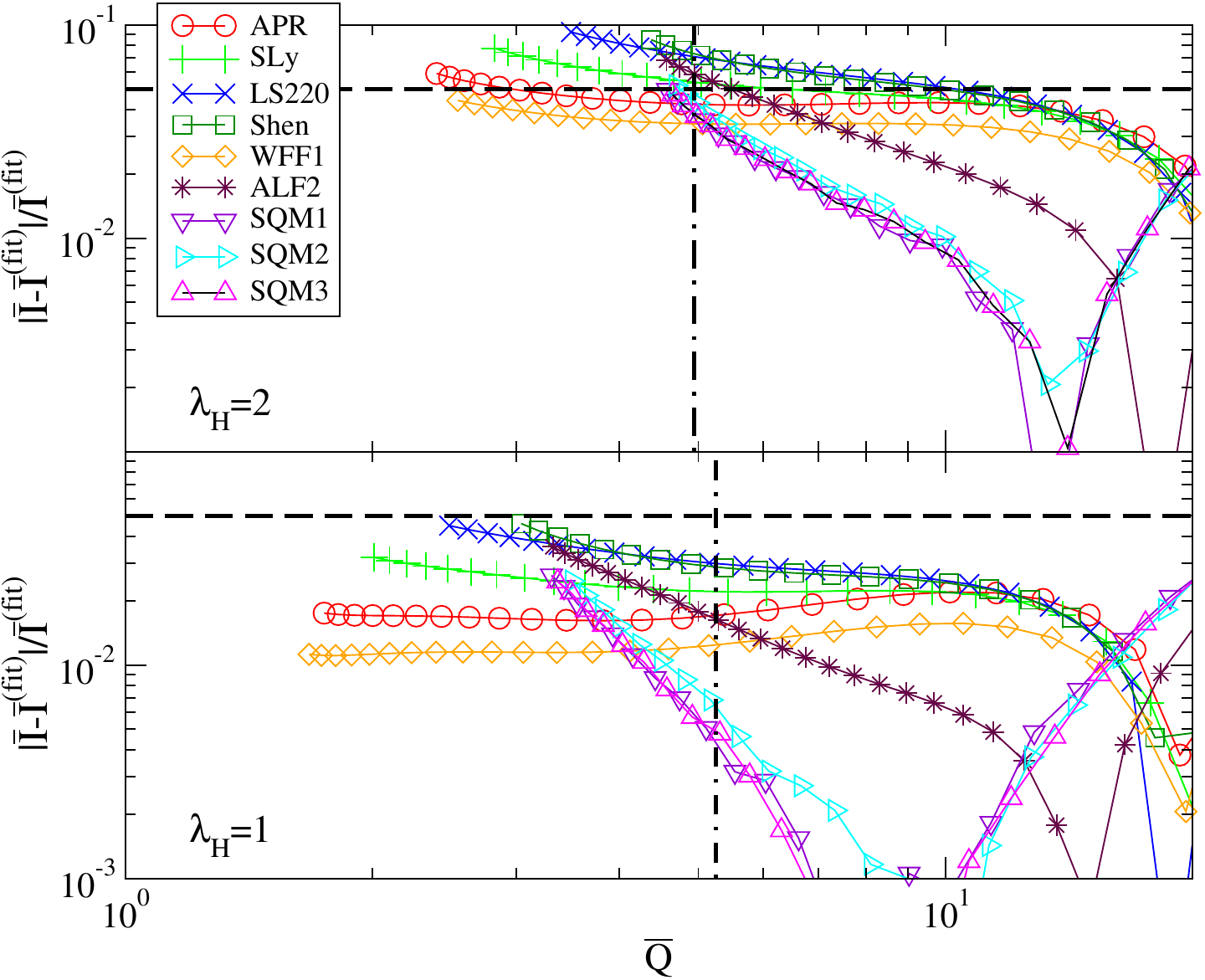}  
\caption{\label{fig:I-Q-diff} (Color online) Fractional difference of the I-Q relation for anisotropic stars ($\lambda_\HH = 2$  in the top and $\lambda_\HH = 1$ in the bottom) with various EoSs relative to the fitting function constructed in~\cite{I-Love-Q-Science,I-Love-Q-PRD} for isotropic stars. A fractional difference below the horizontal dashed line at 5\% is required for NICER and LOFT to achieve the goal of measuring the NS mass and radius to $5\%$ accuracy. The vertical dotted-dashed line shows the value of $\bar Q$ that corresponds to a $M_* = 1.4M_\odot$ NS using an APR EoS with each $\lambda_\HH$. Observe that the fractional error is smaller than $5\%$ if $|\lambda_\HH | \lesssim 1$.
}
\end{center}
\end{figure}

\begin{figure*}[htb]
\begin{center}
\includegraphics[width=8.5cm,clip=true]{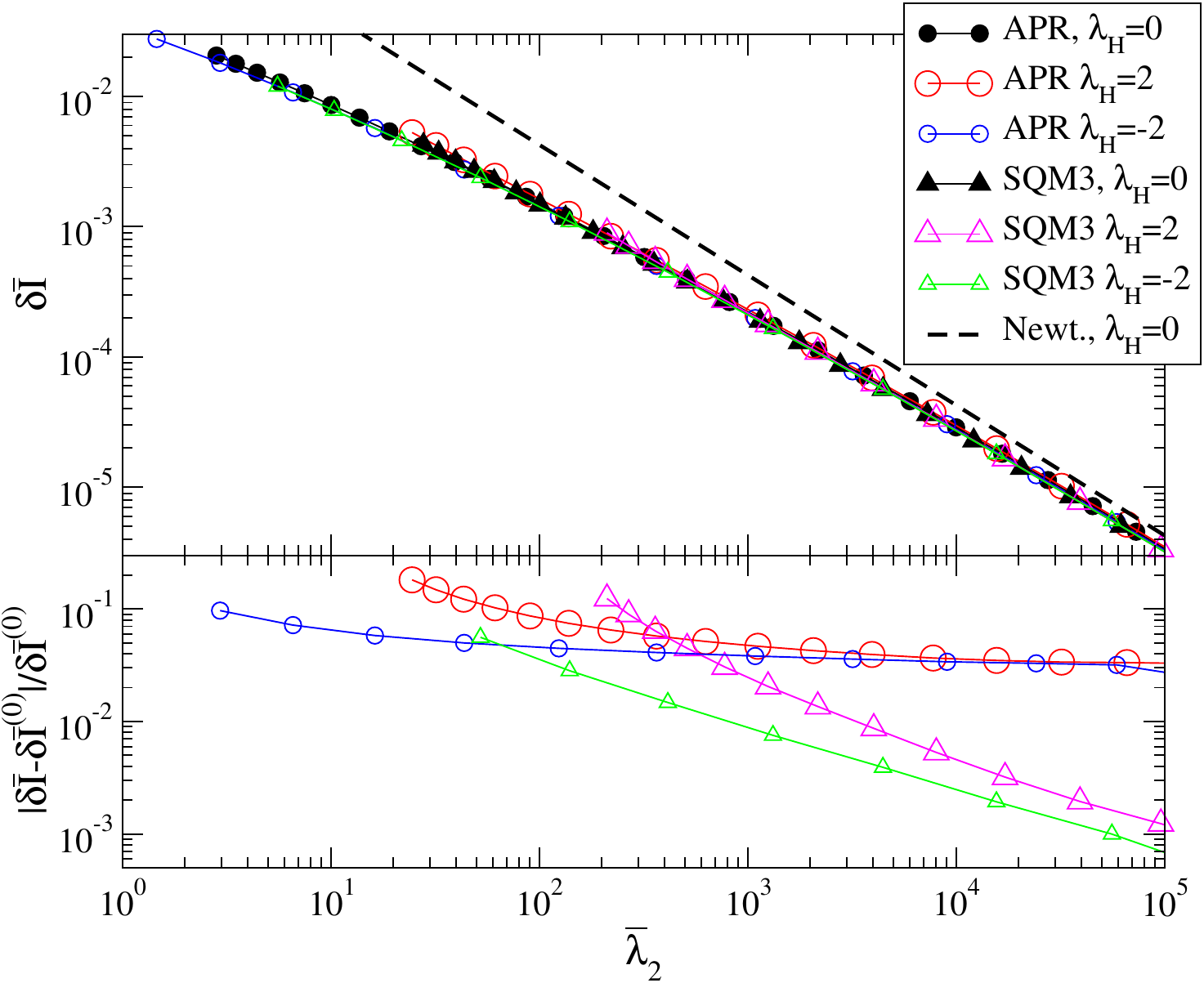}  
\includegraphics[width=8.5cm,clip=true]{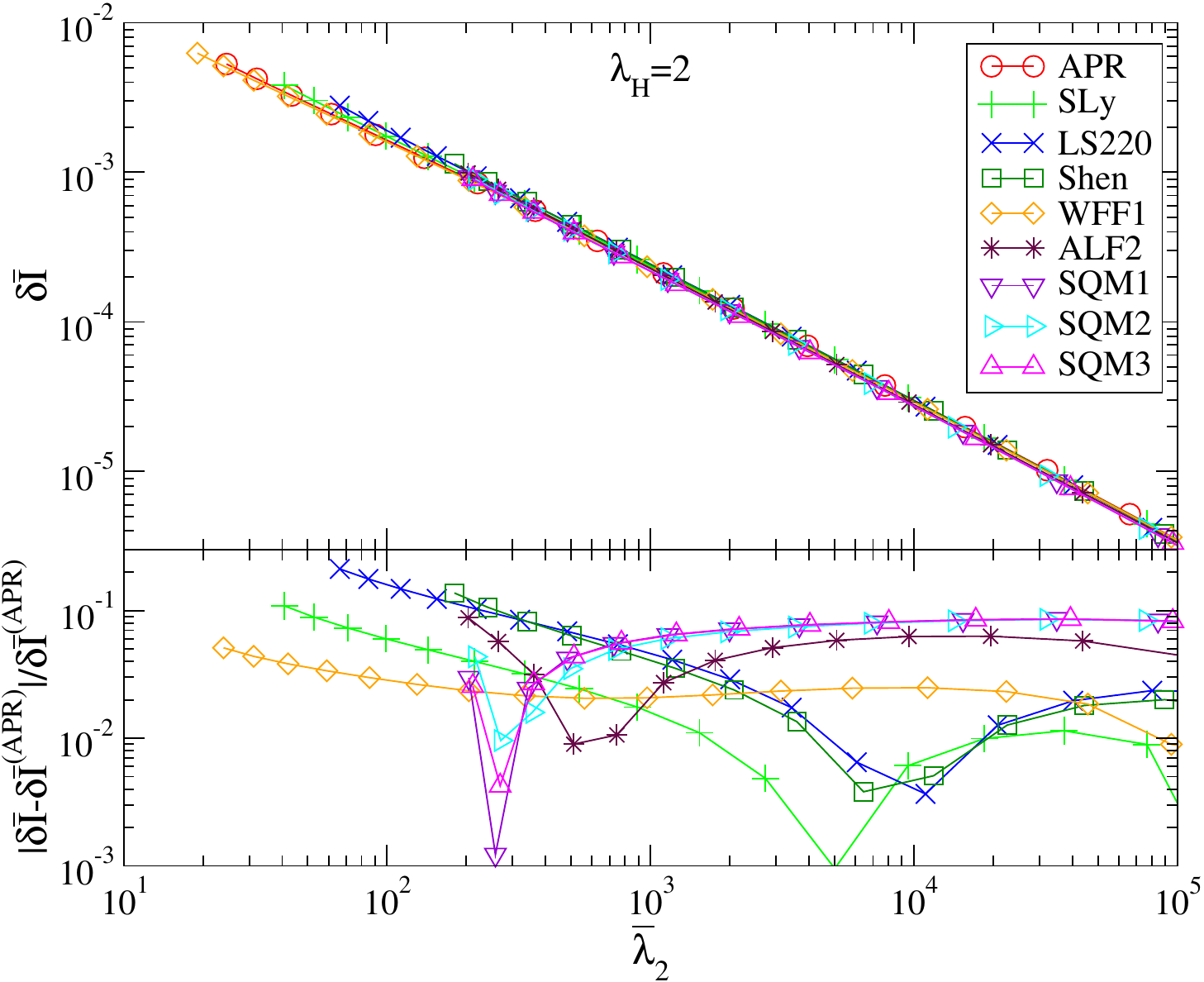}  
\caption{\label{fig:I-Love-CS} (Color online) (Top left) DCS corrections to $\bar I$ as a function of $\bar \lambda_2$ for various anisotropy parameters with an APR and SQM3 EoS.  We also show the analytic relation for an isotropic, $n=0$ polytrope in the Newtonian limit~\cite{alihaimoud-chen}.  (Top right) Same as the top left panel but with $\lambda_\HH=2$ and for various EoSs. (Bottom left) Fractional difference of each relation for anisotropic stars from that of isotropic ones. (Bottom right) Fractional difference of each relation from that with an APR EoS. Observe that the dCS correction to the I-Love relation is both anisotropy and EoS insensitive to $\sim 20\%$ accuracy.
}
\end{center}
\end{figure*}

Let us now study the effect of anisotropy on the application of the I-Q relation to X-ray observations. The goal of NICER~\cite{2012SPIE.8443E..13G} and LOFT~\cite{2012AAS...21924906R,2012SPIE.8443E..2DF,2014SPIE.9144E..2TF} is to measure the NS mass and radius within $5\%$ accuracy through the observations of X-ray pulse profiles from hot spots on the surface of millisecond pulsars. Since a millisecond pulsar is rotating moderately fast, the profile depends not only on the NS mass and radius, but also on other quantities such as the NS moment of inertia and quadrupole moment~\cite{Psaltis:2013zja}. Reference~\cite{Psaltis:2013fha} showed that one can use the I-Q relation and other universal relations found in~\cite{baubock} to break the degeneracy among some of the parameters on the X-ray pulse profile observations so that one can measure the NS mass and radius to a high accuracy. 

We now estimate the amount of anisotropy required such that the I-Q relation can be still used in future X-ray observations even for anisotropic compact stars. Since the measurement accuracy goal of the mass and radius with NICER and LOFT is $5\%$, one needs the anisotropy and EoS variation in the I-Q relation to be smaller than $5\%$ so that the systematic errors do not dominate the statistical errors. From Fig.~\ref{fig:max-diff}, one sees that for fixed $\lambda_\HH$, the EoS-variation on the I-Q relation is $\leq 5\%$. Therefore, one might think that one can create a fit for each $\lambda_\HH$ and apply that to future observations. However, in practice, this cannot be done because we do not know what $\lambda_\HH$ is for each star.  

Let us now study the EoS variation of the I-Q relation for anisotropic stars relative to the fit of the I-Q relation for isotropic stars. The top panel of Fig.~\ref{fig:I-Q-diff} presents the fractional difference of the I-Q relation for anisotropic compact stars ($\lambda_{\HH} = 2$) with various EoSs relative to the fitting formula constructed for the I-Q relation of isotropic stars in~\cite{I-Love-Q-Science,I-Love-Q-PRD}. For reference, we show the 5\% error line and the $M_*=1.4M_\odot$ line as a horizontal dashed and a vertical dotted-dashed curve respectively. Observe that the fractional error exceeds $5\%$, and hence, one is dominated by systematic error if one were to use the fitted relation in~\cite{I-Love-Q-Science,I-Love-Q-PRD} for anisotropic stars with $\lambda_\HH = 2$. The bottom panel of Fig.~\ref{fig:I-Q-diff} shows the same information as the top panel, but for anisotropic stars with $\lambda_\HH=1$. Observe that in this case, the fractional error is below 5\% for all EoS. Therefore, we conclude that one can use the relation for isotropic stars in future X-ray observations if $|\lambda_\HH | \lesssim 1$. Such a value of $\lambda_{\HH}$ corresponds to an anisotropic effect that modifies the moment of inertia and the quadrupole moment by $\leq 10\%$ relative to their values for isotropic stars (see Fig.~\ref{fig:diff-fixed-mass}).

\subsubsection{Experimental Relativity}

Let us conclude with a short discussion of the effect of anisotropy on the application of the I-Love relations when performing strong-field tests of gravity. Although compact stars offer an excellent testbed to probe strong-field gravity, such tests are limited by the degeneracies between uncertainties in nuclear physics and the effect of modified gravity theories. One can break such degeneracies by independently measuring two observables whose interrelation is approximately universal. Such a relation, in principle, will depend only on the underlying gravitational theory, while remaining approximately insensitive to the EoS. 

Among various universal relations, the I-Love relation might be the best for experimental relativity. This is because, in the future, one may be able to measure the moment of inertia from double binary pulsars and the tidal deformability from GW observations. The NSs observed may be different, and in particular, may have different masses, but even so, one can still do such tests, as discussed in~\cite{I-Love-Q-PRD}. Figures~\ref{fig:univ-APR-SQM3-intro} and~\ref{fig:univ-lambda2-2} show that the anisotropy and EoS variations on the I-Love relation for anisotropic stars are $\sim 3\%$. Since this variation is much smaller than the expected measurement accuracy of the moment of inertia and tidal deformability, one concludes that one can safely perform a consistency test of GR even if the star were anisotropic. 

Just because one can perform a consistency test does not necessarily mean that the constraint one would derive remains unaffected by pressure anisotropy. Indeed, pressure anisotropy may have a strong effect on the approximately universal relations in modified gravity. In order to study this in more detail, let us consider the I-Love relation in dCS gravity~\cite{jackiw,CSreview} as an example. Such a theory, motivated by string theory and loop quantum gravity, is a modification to GR that introduces a parity-violating scalar field that couples non-minimally to a quadratic curvature scalar (see App.~\ref{sec:DCS-ABC} for a more detailed description of the theory.) DCS gravity has a characteristic length scale $\xi_\CS^{1/4}$ that controls the magnitude of any deviation from a GR prediction. The current most stringent bound on this quantity is $\xi_\CS^{1/4} < \mathcal{O}(10^8)$km, which comes from Solar System and table-top experiments~\cite{alihaimoud-chen,kent-CSBH}.

In order to investigate the effect of anisotropy on the I-Love relation in dCS gravity, we first construct slowly-rotating anisotropic stars in this theory, following Sec.~\ref{sec:DCS-formalism}. The tidal deformability is the same as that in GR due to parity considerations~\cite{quadratic}, but the moment of inertia is dCS modified.  We define the fractional dCS correction to the moment of inertia by
\be
\delta \bar I = \frac{M_*^4}{\xi_\CS} \frac{\bar I^\CS}{\bar I^\GR}\,,
\ee
where $\bar I^\GR$ and  $\bar I^\CS$ correspond to the GR and dCS contributions to $\bar I$ respectively. Since $\bar I^\CS$ is linearly proportional to $\xi_\CS$, $\delta \bar I$ is independent of $\xi_\CS$.

The top left panel of Fig.~\ref{fig:I-Love-CS} shows the dCS correction to the I-Love relation for compact stars with an APR and SQM3 EoS and various fixed values of $\lambda_\HH$. The bottom left panel of  Fig.~\ref{fig:I-Love-CS} shows the fractional difference of each I-Love relation relative to the isotropic result. The right panel shows the I-Love relation for various EoSs (top) and its fractional difference with respect to the APR EoS (bottom), all for anisotropic stars with a fixed $\lambda_\HH = 2$. Observe that the anisotropy and EoS variations in the dCS I-Love relation is $\sim 20\%$, which is an order of magnitude larger than that in GR. In the top left panel, we also show the dCS correction to the I-Love relation in the Newtonian limit for incompressible, isotropic star, which we derive from the solution obtained in~\cite{alihaimoud-chen}: $\delta \bar I  = 32/75 \bar \lambda_2^{-1}$. Observe that numerical calculations for anisotropic stars approach this isotropic Newtonian relation as one increases $\bar \lambda_2$. We expect this is because anisotropy vanishes in the H model in the Newtonian limit.

Clearly, anisotropy affects the I-Love relation a lot in dCS gravity; in fact, it does so much more than in GR. But how does such an increased variation due to anisotropy in dCS affect the use of the I-Love relation in strong-field tests of GR. Following~\cite{I-Love-Q-Science,I-Love-Q-PRD}, we assume that one can measure the moment of inertia to 10\% accuracy from a future double binary pulsar observations and the tidal deformability to 40\% accuracy with future GW observations. Assuming GR is the correct theory, one can draw a fiducial point on the GR I-Love relation with an error box around it in the I-Love plane, as shown in Fig.~\ref{fig:I-Love-Shen-error}. We also show here the dCS I-Love relation with $\xi_\CS / M_*^4 = 1.85 \times 10^4$, which is the value we chose in~\cite{I-Love-Q-Science,I-Love-Q-PRD} so that the relation is marginally consistent with this hypothetical measurement for isotropic NSs, i.e.~this is the largest value of $\xi_{\CS}$ for which the I-Love dCS curve still crosses the error box. 

Several points need to be made about the results shown in Fig.~\ref{fig:I-Love-Shen-error}. First, observe that the constraint on $\xi_{\CS}$ when assuming isotropic stars is the same as that obtained when assuming anisotropic stars with $\lambda_\HH = -2$. However, when $\lambda_{\HH} = 2$ this is not the case anymore. This is because the dCS I-Love curve when $\lambda_{\HH} = 2$ shifts upwards, which then allows for the more stringent constraint of $\xi_\CS / M_*^4 \leq 1.80 \times 10^4$. Notice that such a bound differs from the isotropic one by only $\sim 3\%$, which does not affect at all the fact that such a bound is still six orders of magnitude stronger than the current bound. Moreover, since we do not know whether NSs are anisotropic, and if they are, whether anisotropy plays such a strong role as in the case when $\lambda_{\HH} = 2$, the conservative choice is to assume the weaker bound $\xi_\CS / M_*^4 \leq 1.85 \times 10^4$, derived from the isotropic analysis. 

\section{Conclusions and Discussions}
\label{sec:conclusions}

We have investigated the effect of pressure anisotropy on certain EoS-insensitive, approximately universal relations among observables associated with NSs and QSs. We extended the Hartle-Thorne approach~\cite{hartle1967,Hartle:1968ht} for isotropic stars to anisotropic stars valid to third order in a small spin period expansion. We found that anisotropy increases the variability of the approximately universal relations, for example by a factor of 2--4 relative to the isotropic case when considering the maximum amount of anisotropy allowed by our model. Astrophysically realistic amounts of anisotropy would lead to an insignificant effect on the universal relations.  

Next, we extended the analysis of~\cite{Stein:2014wpa} and derived three-hair relations among multipole moments for anisotropic stars in the non-relativistic limit using polytropic EoS and the elliptical isodensity approximation. We found that the relations are insensitive to anisotropy for polytropes with an index close to $n=0$. This finding confirms analytically the small dependence on anisotropy in the S$_3$-Q relation. We also looked at the stellar eccentricity profile of anisotropic stars and found that the eccentricity variation reaches close to $40\%$. Such a variation can explain the $\sim 10\%$ variation in the S$_3$-Q relation, giving further support to the model of~\cite{Yagi:2014qua} that suggests the origin of the universality as an emergent approximate symmetry related to self-similarity in isodensity contours.  

We then studied how the anisotropy effect affects the applications of universal relations to future observations. This, of course, ultimately depends on the amount of anisotropy inside NSs, a quantity that is currently unknown. Our conclusions apply for the choice of the anisotropy range we considered, which is the same as that chosen in previous work~\cite{Doneva:2012rd,Silva:2014fca}, and it is based on anisotropy due to crystallization of the stellar core~\cite{Nelmes:2012uf} and pion condensation~\cite{Sawyer:1972cq}. Regarding GW astrophysics, we concluded that one does not need to worry about anisotropy if one measures the NS spins to $\sim 10\%$ accuracy. Regarding X-ray nuclear astrophysics, we found that one can use the I-Q relations for isotropic stars if $|\lambda_\HH | < 1$. Regarding experimental relativity, we found that the consistency tests of GR with the I-Love relation are essentially unaffected by anisotropy. As an example, we constructed slowly-rotating anisotropic compact star solutions to linear order in spin in dCS gravity and derived the dCS correction to the I-Love relation. We found that future observations could still place a constraint that is roughly six orders of magnitude stronger than the current bound even if pressure anisotropy were present in NSs.

Possible avenues for future research include extending the analysis presented here to higher order in spin~\cite{Yagi:2014bxa} or even to a rapid rotation, using e.g.~the RNS~\cite{stergioulas_friedman1995} or LORENE~\cite{bonazzola_gsm1993,bonazzola_gm1998} open source numerical codes. One could then study how anisotropy affects the relations between higher multipole moments. Another avenue for future research is to consider anisotropy models other than the H and BL models, and see if the effect of anisotropy is still small. One can also look at the effect of pressure anisotropy on other universal relation, such as the multipole Love relations~\cite{Yagi:2013sva}, and see how it affects the applications of such relations to future observations. Finally, we here considered dCS gravity as an example of a modified gravity theory when considering tests of GR, but one could also study the universal relations for anisotropic stars in other modified theories of gravity.

{\emph{Acknowledgments}}.~We would like to thank Jim Lattimer and Bennett Link for useful comments, suggestions and advice. N.Y. acknowledges support from NSF grant PHY-1114374, NSF CAREER Grant PHY-1250636 and NASA grant NNX11AI49G. Some calculations used the computer algebra-systems MAPLE, in combination with the GRTENSORII package~\cite{grtensor}.

\appendix

\section{Slowly-rotating Anisotropic Stars in dCS Gravity}
\label{sec:DCS}

In this appendix, we present the basics of dCS gravity and how to construct slowly-rotating, anisotropic compact stars in this theory. Since the focus of this paper, when referring to dCS gravity, is on the I-Love relation, and since the tidal deformability in dCS is the same as that in GR~\cite{quadratic,I-Love-Q-PRD}, we only consider slowly-rotating NS solutions valid to linear order in spin.

\subsection{ABC of dCS Gravity}
\label{sec:DCS-ABC}

The action for dCS gravity is given by~\cite{CSreview}
\begin{align}
S &\equiv  \int d^4x \sqrt{-g} \Big[ \kappa_g R + \frac{\alpha}{4} \vartheta R_{\nu\mu \rho \sigma} {}^* R^{\mu\nu\rho\sigma}  \nn \\
&   - \frac{\beta}{2} \nabla_\mu \vartheta \nabla^{\mu} \vartheta + \mathcal{L}_{\MAT} \Big]\,.
\label{action}
\end{align}
Here, $\kappa_g \equiv (16\pi)^{-1}$, $g$ represents the determinant of the metric $g_{\mu\nu}$ and $R_{\mu\nu \rho \sigma}$ represents the Riemann tensor. ${}^* R^{\mu\nu\rho\sigma}$ is its dual~\cite{CSreview}, $\vartheta$ is a scalar field (not to be confused with the LE function $\vartheta_\LE$) while $\alpha$ and $\beta$ are coupling constants and $\mathcal{L}_{\MAT}$ is the matter Lagrangian density. We neglect the potential of the scalar field for simplicity~\cite{yunespretorius,quadratic,kent-CSBH}. We take $\vartheta$ and $\beta$ to be dimensionless, which means $\alpha$ must have dimensions of (length)$^2$~\cite{yunespretorius}.
We define the dimensionless coupling parameter $\zeta_\CS$ as
\be
\zeta_\CS \equiv \frac{\xi_\CS M_*^2}{R_*^6}\,,
\quad \xi_\CS \equiv \frac{\alpha^2}{\kappa_g \beta}\,.
\ee
We treat dCS gravity as an effective theory and keep terms up to linear order in $\zeta_\CS$ in all equations. Otherwise, the theory is ill-posed~\cite{Delsate:2014hba} without the inclusion of higher-order curvature terms.

The field equations in dynamical CS gravity are given by~\cite{CSreview}
\be
G_{\mu\nu} + \frac{\alpha}{\kappa_g} C_{\mu\nu} =\frac{1}{2\kappa_g} (T_{\mu\nu} + T_{\mu\nu}^\vartheta)\,,
\label{eq:field-eq-CS}
\ee
where $G_{\mu\nu}$ is the Einstein tensor. The C-tensor and the stress-energy tensor for the scalar field are defined by
\begin{align}
C^{\mu\nu} & \equiv  (\nabla_\sigma \vartheta) \epsilon^{\sigma\delta\alpha(\mu} \nabla_\alpha R^{\nu)}{}_\delta + (\nabla_\sigma \nabla_\delta \vartheta) {}^* R^{\delta (\mu\nu) \sigma}\,, \\
\label{eq:Tab-theta}
T_{\mu\nu}^\vartheta & \equiv  \beta (\nabla_\mu \vartheta) (\nabla_\nu \vartheta) -\frac{\beta}{2} g_{\mu\nu} \nabla_\delta \vartheta \nabla^\delta \vartheta\,.
\end{align}
On the other hand, the evolution equation for the scalar field is given by
\be
\square \vartheta = -\frac{\alpha}{4 \beta} R_{\nu\mu \rho \sigma} {}^*R^{\mu\nu\rho\sigma}\,.
\label{scalar-wave-eq}
\ee
%

\subsection{Field Equations for Slowly-Rotating  \\ Anisotropic Stars}
\label{sec:DCS-formalism}

We now explain how one can extend the GR analysis explained in the main text to dCS gravity and construct slowly-rotating, anisotropic relativistic stars to linear order in a small spin period expansion. 
Let us first look at the scalar field equation. To do so, we first introduce a book keeping parameter $\alpha'$ to count factors of $\zeta_{\CS}^{1/2}$. We are only interested in studying the scalar field to $\mathcal{O}(\alpha')$, since higher-order corrections in $\alpha'$ only enter at $\mathcal{O}(\alpha'{}^3)$ in the metric, which is higher than $\mathcal{O}(\zeta_\CS) = \mathcal{O}(\alpha'^2)$. Since the leading-order contribution to $\vartheta$ is $\mathcal{O}(\alpha')$ and the right-hand-side of Eq.~\eqref{scalar-wave-eq} has a factor of $\alpha$, one only needs to consider the GR contribution to the metric. We decompose the scalar field $\vartheta$ as~\cite{yunes-CSNS,alihaimoud-chen,Yagi:2013mbt} 
\be
\label{eq:scalar-decomp}
\vartheta = \sum_{\ell=0} \vartheta_\ell (R) P_\ell (\cos \theta)\,,
\ee
and substitute this, together with the metric ansatz in GR in Hartle-Thorne coordinates, into Eq.~\eqref{scalar-wave-eq}. Then, one finds that the only equation with a non-vanishing source term is the $\ell =1$ mode, given by
\begin{align}
& \frac{d^2 \vartheta_1}{d R^2} + \frac{1 + e^\lambda \left[ 1 -4\pi R^2 (\rho-p)  \right]}{R} \frac{d \vartheta_1}{d R}  - 2\frac{e^\lambda}{R^2} \vartheta_1 \nn \\
&=  16\pi \frac{\alpha}{\beta}  (\delta - \sigma_0) e^{(\lambda - \nu)/2} \frac{d \omega_1}{d R}\,, 
\label{vartheta1RR}
\end{align}
with 
\be
\delta \equiv \rho- \frac{3 M}{4 \pi R^3}\,.
\ee

Let us now look at the evolution equation for the metric perturbation to linear order in spin. We note that spherically symmetric spacetimes (non-rotating stars) are unaffected in dCS due to parity considerations. We thus start by decomposing $\omega_1(R)$ as
\be
\label{eq:omega1-decomp-CS}
\omega_1 (R) = \omega^{\GR}_{1} (R) - \alpha' \omega_1^\CS (R)\,,
\ee
where $ \omega^{\GR}_{1}$ and $ \omega^{\CS}_{1}$ represent the GR and dCS contributions to $\omega_1$ respectively. We introduced a minus sign in front of $\alpha' \omega^{\CS}_{1}$ so that the CS correction to $\omega_1$ matches the conventions in~\cite{Yagi:2013mbt}. Next, we substitute the metric ansatz and Eqs.~\eqref{eq:scalar-decomp} and~\eqref{eq:omega1-decomp-CS} into the field equations in Eq.~\eqref{eq:field-eq-CS}. At $\mathcal{O}(\epsilon \; \alpha')$, one finds
\begin{align}
& \frac{d^2 \omega^{\CS}_{1}}{d R^2} + 4 \frac{1-\pi R^2 (\rho +p) e^\lambda}{R} \frac{d \omega^{\CS}_{1}}{d R} 
-16 \pi (\rho +p - \sigma_0) e^\lambda \omega^{\CS}_{1} \nn \\
&= - \frac{128 \pi^2 \alpha e^{(\nu+\lambda)/2}}{R^3} \left[ \delta \; (R-\sigma_0) \frac{d \vartheta_1}{d R} 
\right. \nn \\ 
& \left. 
+ \left( R \frac{d\rho}{dR} - \delta - 2 \sigma_0 - \sigma'_0R \right) \vartheta_1 \right]\,. 
\end{align}

In order to obtain the interior solution for $\vartheta_1$ and $\omega^{\CS}_{1}$, one needs a boundary condition at the stellar center and at the surface. The asymptotic behavior of the scalar field at the former is given by
\be
\vartheta_1 (R) = \vartheta'_c r_\epsilon + \frac{2\pi}{15} (5\rho_c - 3p_c) \vartheta'_c r_\epsilon^3 + \mathcal{O}(x^4)\,,
\label{vartheta0}
\ee
in both the H and BL models, with $\vartheta'_c$ representing a constant that corresponds to $d\vartheta/dR$ at the center and we recall that $x = r_\epsilon/R_* \ll 1$. On the other hand, the asymptotic behavior of $\omega^{\CS}_{1}$ at the stellar center for the H and BL models is different and given by
\begin{align}
\allowdisplaybreaks
\omega^{\CS,\HH}_{1} (r_\epsilon) &= \omega_{1c}^{\CS} + \frac{8\pi}{15} \left[ 3 (\rho_c + p_c) \omega_{1c}^\CS  -16\pi \alpha e^{\nu_c/2} (3 \rho_2 \right. \nn \\
&-  \left.  20 \pi  \lambda_H p_c \rho_c) \vartheta'_c \right] r_\epsilon^2 + \mathcal{O}(x^3)\,, \\
\omega^{\CS,\BL}_{1} (r_\epsilon) &= \omega_{1c}^\CS + \frac{8 \pi}{15} \left\{ 3 \omega_{1c}^\CS (\rho_c + p_c)  - 8 \pi \alpha  e^{\nu_c/2}  [ 6 \rho_2 \right. \nn \\
&-  \left.  5 \lambda_\BL  (p_c + \rho_c) (3 p_c+ \rho_c) ]  \vartheta'_c  \right\}r_\epsilon^2 + \mathcal{O}(x^3)\,, \nn \\
\label{w0-CS}
\end{align}
respectively, where $ \omega_{1c}^{\CS} = \omega^{\CS}_{1} (0)$ is a constant that needs to be determined through matching.

The boundary condition at the surface is obtained as follows. The exterior solution for $\vartheta_1$ and $\omega^{\CS}_{1}$ are given in~\cite{alihaimoud-chen,Yagi:2013mbt}. In particular, the asymptotic behavior of $\omega^{\CS}_{1}$ at spatial infinity is given by
\be
\omega^{\CS}_{1} (R) = 2 \bar I^{\CS} \Omega \frac{M_*^3}{R^3} + \mathcal{O}\left( \frac{M_*^4}{R^4} \right)\,.
\ee
Here, $\bar I^\CS$ corresponds to the dCS correction to $\bar I$. Again, $\bar I^\CS$ is determined through the matching at the surface. Such a boundary condition is given by Eq.~\eqref{eq:matching} with $A = \vartheta_1$ and $\omega^{\CS}_{1}$. For a star with a discontinuous density at the surface, the matching condition for $\omega^{\CS}_{1}{}'$ is given by Eq.~\eqref{eq:matching-jump} with 
\be
j_{\omega^{\CS}_{1}} =  \frac{128 \pi^2 \alpha e^{(\nu_* + \lambda_* )/2}  \vartheta_{1,*}}{R_*^2}\,.
\ee
%

\bibliography{master}
\end{document}